\documentclass[graybox]{svmult}

\usepackage{type1cm}        
\usepackage{makeidx}         
\usepackage{graphicx}        
\usepackage{multicol}        
\usepackage[bottom]{footmisc}

\usepackage{newtxtext}       %
\usepackage{newtxmath}       
\usepackage{verbatim}
\usepackage{subfigure}
\usepackage{times}
\usepackage{caption}
\usepackage{epsfig}
\usepackage{stfloats}
\usepackage[noadjust]{cite}
\usepackage{enumerate}
\usepackage{multirow}
\usepackage{latexsym}
\usepackage{color}
\usepackage{amsmath}
\usepackage[noend]{algpseudocode}
\usepackage{epstopdf}
\usepackage{tikz}
\usepackage[american]{circuitikz}
\usepackage{algorithm}
\usepackage{mathtools}

\makeindex             

\begin{document}

\title*{Optimization of Resources to Minimize Power Dissipation in 5G Wireless Networks}
\author{Jyotsna Rani and Ganesh Prasad}
\institute{Jyotsna Rani \at \email{jyotsnarani0009@gmail.com}
\and Ganesh Prasad \at Dept. of Electronics and Communication Engineering, National Institute of Technology Silchar, \email{gpkeshri@ece.nits.ac.in}}

%
\maketitle

\abstract*{In today’s modern communications, with evolution of various applications, the demand of data rate is increasing exponentially at the cost of huge consumption of available resources. It has been recorded that the communication networks dissipate nearly 1\% of the world-wide total power consumption, results in millions of tons of $CO_2$ emission due to their production and thereby causes various environmental health hazards. The optimal utilization of available resources that can balance the present coexisting problem without any compromise on the high throughput demand, paves the way for the next generation green 5G wireless networks. In this chapter, we study the minimization of total power consumption while satisfying the desired coverage of the user equipments (UEs) to provide the minimum throughput over the network. In this regard, the deployment of base stations (BSs), their number, and transmit power are optimized in two scenarios (i) when the UEs are large in 5G wireless network and (ii) when moderate UEs are distributed over the field.}

\abstract{In today’s modern communications, with evolution of various applications, the demand of data rate is increasing exponentially at the cost of huge consumption of available resources. It has been recorded that the communication networks dissipate nearly 1\% of the world-wide total power consumption, results in millions of tons of $CO_2$ emission due to their production and thereby causes various environmental health hazards. The optimal utilization of available resources that can balance the present coexisting problem without any compromise on the high throughput demand, paves the way for the next generation green 5G wireless networks. In this chapter, we study the minimization of total power consumption while satisfying the desired coverage of the user equipments (UEs) to provide the minimum throughput over the network. In this regard, the deployment of base stations (BSs), their number, and transmit power are optimized in two scenarios: (i) when the UEs are large in 5G wireless network and (ii) when moderate UEs are distributed over the field.}

\section{Introduction and Background}\label{sec:introduction}
Today, the digital domain of all applications has become an integral part of our daily life that results in the growth of data rate demand by 10 times in every 5 years. In order to fulfill it, the architecture is also increasing at the same rate in the next generation wireless network that consumes around 1\% of the world-wide total electricity consumption, it requires a huge production of electricity which emits 130 million tons of $CO_2$ per year~\cite{fett01}. Therefore, we need a reasonable strategy for optimal utilization of available resources to combat the coexistence problem while satisfying the desired coverage of UEs. 

In this chapter, we investigate the optimal deployment of network architecture as well as the optimization of transmit power to minimize overall power consumption under the given constraint of desired coverage that gives the minimum throughput over the network. From the state-of-the-art, authors in~\cite{andr02,elsa03,srin04} have deployed the network nodes randomly with a given distribution. It has been described in~\cite{andr02} that deployment based on homogeneous Poisson point process (HPPP) is more tractable and satisfies the practical aspects than the conventional strategy where the nodes are uniformly placed on a grid. In~\cite{elsa03}, for different types of network and MAC layer, various point processes like HPPP, Binomial point process (BPP), hard core point process (HCPP), and Poisson cluster process (PCP) are used to measure the system performances. Further, it is analyzed in~\cite{srin04} that BPP is a more realistic and tractable model than HPPP.

 Based on the required hours for service of the BSs, the power dissipation over the network can be reduced by dynamically turn them on/off~\cite{wua11}. Different frameworks used to determine the sleeping mode are discoursed in~\cite{Ajm03,oh05}. In~\cite{Ajm03}, the on/off the BSs is determined by the traffic profile, besides, traffic profile as well as BSs density are used for deciding the sleeping mode in~\cite{oh05}. Energy saving algorithm based on on/off of the BSs that achieves saving upto $80\%$ is discoursed in~\cite{oh06}. The tradeoff between energy efficiency and spectral efficiency is investigated in~\cite{peng07} over a switching based network, thereafter, a power control technique is described to optimize the tradeoff. Authors in~\cite{zhou08} describe the centralized and decentralized power reduction mechanism while satisfying the outage constraint. The blockage in services and delay in delivery of the data due to switching operation are analyzed in~\cite{jie09,wua10}.

Recently, the optimization of deployment and transmit power have been investigated in~\cite{per05,sar06,ver07,gon08}. In~\cite{per05,sar06}, the area power consumption of the network is minimized by optimizing the density of the base stations (BSs) under the constraint of users' coverage and data rate demand. Authors in~\cite{ver07} optimize the multiple parameters like transmit power, density of BSs, number of antennas and users served per BS to reduce the overall power dissipation while satisfying a given data rate demand. A different strategy for energy saving is described in~\cite{gon08} where the number of BSs and their location are optimized. However, from the recent works, the optimization of transmit power, number of nodes and their location is further required to be explored in two scenarios that are: (i) when large UEs are associated in a 5G wireless network and (ii) when moderate UEs are distributed in a rural area.

\section{System Assumptions}\label{sec:sys_asum}
Here, we describe the network topology and channel models along with system assumptions considered for the proposed framework.

\begin{figure}[!t]
	\sidecaption
	\includegraphics[scale=.13]{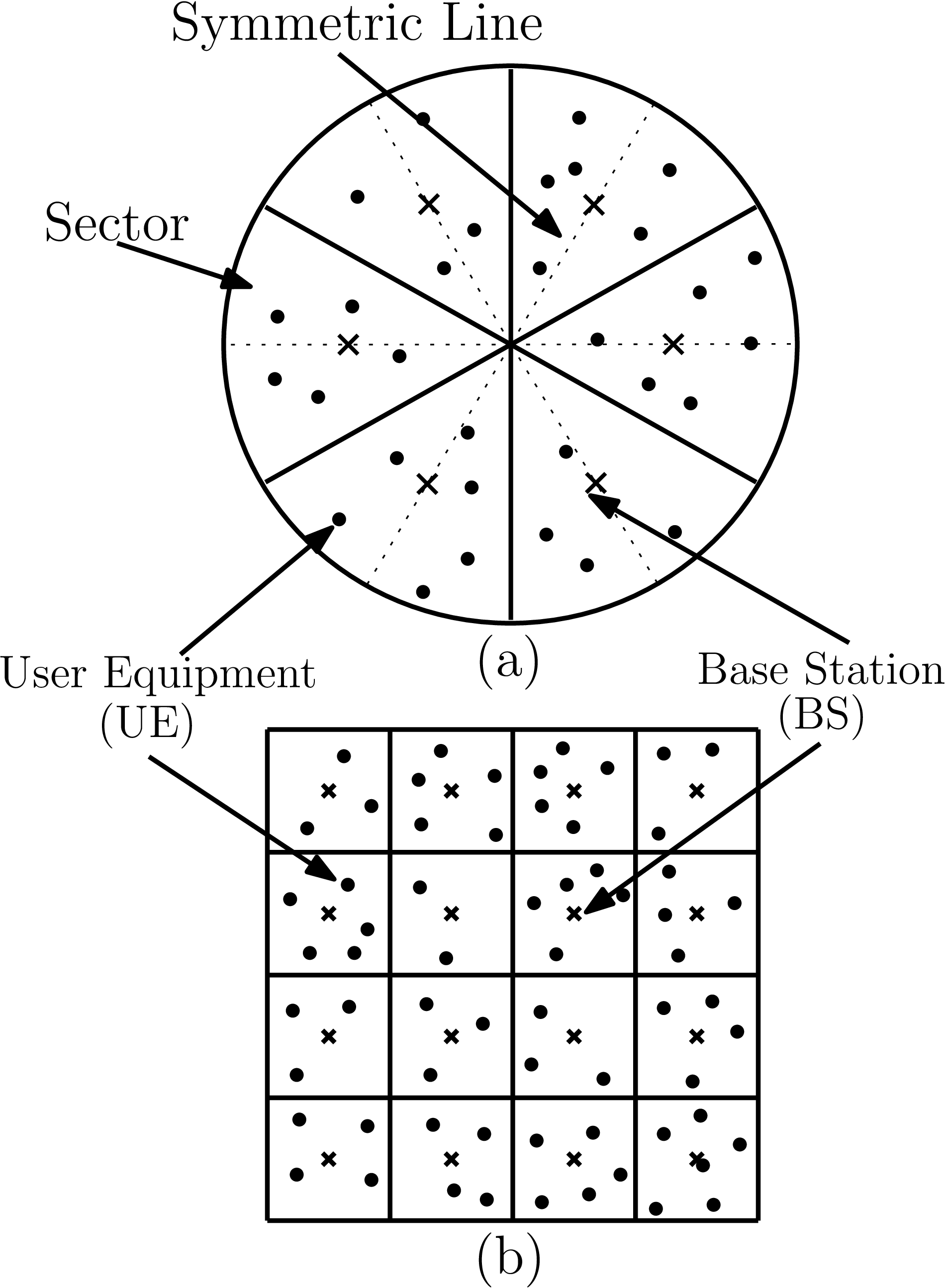}
	\caption{\small Randomly distributed UEs by BPP lying under the coverage of deterministically deployed BSs in (a) a circular and (b) a square field.}
	\label{fig:network_topo}
\end{figure}  

\begin{figure}[!t]
	\sidecaption
	\includegraphics[scale=0.4]{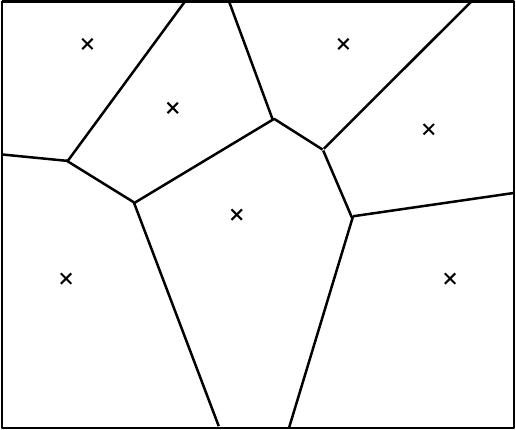}
	\caption{\small Cells generations using Voronoi tessellation about the BSs (represented by $\times$) in a bounded square field.}
	\label{fig:voronoi}
\end{figure}

\subsection{Network Topology}\label{sec:nwk_topo}
We consider a circular and a square field where the $N_u$ UEs are distributed using BPP and $N_b$ BSs are deterministically deployed as shown in Fig.~\ref{fig:network_topo}. The size of the cells over the field is determined by Dirichlet regions using Voronoi tessellation~\cite{sto09} as depicted in Fig.~\ref{fig:voronoi}. Here, the boundary of a cell is determined by the bisector line between the two BSs (denoted by $\times$). Based on it, the generated cells in Fig.~\ref{fig:network_topo} are square or rectangular in shape over the square field as the BSs are located at regular distance along the length and width. Similarly, the generated cells over the circular field have the shape of arc or triangular as the BSs are placed at a regular intervals in the angular direction from the center. A UE lying in a particular cell is assumed to be completely associated with the underlying BS in the cell. The interference over the downlink communication is assumed to be absent as BSs use orthogonal multiple-access techniques~\cite{mhi10}.

\subsection{Channel Model}
We assume that the channel is flat where the channel coefficient $h$ between an UE and a BS is a variate with Rayleigh distribution. Therefore, $h^2$ has exponential distribution denoted as $h^2\sim\exp(\mu)$ with mean $\mu=1$.
The signal-to-noise ratio (SNR) $\gamma$ received at a UE from a BS at a distance $r$ is given by $\gamma = \frac{P_t h r^{-\alpha}}{\sigma^2}$, where $P_t$ is the transmit power from the BS, $\alpha$ is the path loss exponent, and $\sigma^2$ is the noise power of the additive noise at the receiver~\cite{rap11}. In the framework, we assume that the network is homogeneous as all the BSs transmit the same power $P_t$.

\subsection{Power Dissipation in a BS}
While a downlink transmission, the power dissipation $P_{_D}$ in a BS is given as:
\begin{align}
	P_{_D}=\upepsilon_1 P_t + \upepsilon_2,
\end{align}
where $\upepsilon_1$ accounts for the scaling of the power transmitted by the BS and $\upepsilon_2$ is power dissipation due to signal processing, power supply, and battery backup. Total power consumption over $N_b$ BSs deployed over the field is $N_b P_{_D}$ that is required to be minimized by optimizing the available resources.

\section{Problem Formulation and Optimization of Transmit Power}\label{Sec:Perform_n_formul}
In this section, we study the coverage probability of a UE by a BS. Thereafter, coverage probability of the farthest UE from the BS is described in two cases: (i) when a single BS and (ii) when multiple BSs are deployed over the field. Lastly, a constrained optimization problem is formulated to minimize the total power consumption $N_b P_{_D}$.

\subsection{Coverage Probability of an UE}
A UE is said to be in coverage from its BS when the SNR $\gamma \geq \mathcal{T}$, where $\mathcal{T}$ is threshold SNR at the UE to successfully detect the received signal. If the UEs are distributed using BPP, coverage probability of a UE from the BS is given as:
\begin{align}\nonumber\label{eq:cov}
P_{cov}&=\mathbb{E}_r \left[\mathrm{Pr}\left(\dfrac{P_thr^{-\alpha}}{\sigma^{2}}\geq \mathcal{T}\right)\right] = \mathbb{E}_r\left[\mathrm{Pr}\left(h\geq\dfrac{\mathcal{T}\sigma^{2}r^\alpha}{P_t}\right)\right]\\
&=\int_r\exp\left(-\dfrac{\mathcal{T}\sigma^{2}r^\alpha}{P_t}\right)f\left(r,x,y\right)\mathrm{d}r,
\end{align}
where  $\mathbb{E}_r[\hspace{0.5mm}\cdot\hspace{0.5mm}]$ is the expectation in distance $r$ of the UE from the BS and $f(r,x,y)$ is probability density function (PDF) of $r$ when the BS is located at $(x,y)$.

\subsection{Coverage Probability of the Farthest UE}
In order to assure the coverage of all UEs, we need to investigate the coverage of the farthest UE from the BS in a cell. This can be analyzed in two cases when a cell has: (i) a single BS and (ii) multiple BSs.

\subsubsection{Case (i)} 
If $N_u$ UEs are distributed using BPP over a cell consisting a single BS, then the cumulative density function (CDF) $F_{far}(r,x,y)$ and PDF $f_{far}(r,x,y)$ of the farthest UE distance from the BS are given as~\cite{tho12}:
\begin{align}\label{eq:CDF_PDF_Far}
	F_{far}(r,x,y)=[F(r,x,y)]^{N_u};\;f_{far}(r,x,y)=N_u[F(r,x,y)]^{N_u-1}f(r,x,y)
\end{align}
Further, using~\eqref{eq:cov}, the coverage probability of farthest UE is given as:
\begin{align}\label{eq:dist_single}
P_{cov}^{far}=
\int_r\exp\left(-\frac{\mathcal{T}\sigma^{2}r^\alpha}{P_t}\right)f_{far}\left(r,x,y\right)\mathrm{d}r.
\end{align}

\subsubsection{Case (ii)}
For $N_u$ UEs distributed using BPP over a cell consisting $N_b$ $\left(<N_u\right)$ BSs, the CDF and PDF of farthest UE distance from the $i^{th}$ BS is given by~\eqref{eq:CDF_multi} and \eqref{eq:PDF_multi}, respectively.
\begin{align}\label{eq:CDF_multi}
	F_{far}^i(r,\mathbf{x},\mathbf{y})&=\sum_{k=0}^{N_u} {N_u \choose k}\left(\frac{A_i}{W}\right)^k\left(1-\frac{A_i}{W}\right)^{N_u-k}\big[F_i(r,\mathbf{x},\mathbf{y})\big]^k,\\\label{eq:PDF_multi}
	f_{far}^i(r,\mathbf{x},\mathbf{y})&=\sum_{k=0}^{N_u} {N_u \choose k}\left(\frac{A_i}{W}\right)^k\left(1-\frac{A_i}{W}\right)^{N_u-k}k\big[F_i(r,\mathbf{x},\mathbf{y})\big]^{k-1}f_i(r,\mathbf{x},\mathbf{y}),
\end{align}
where $W$ is the total area of the field, $A_i$ is the area of the $i^{th}$ cell, $(\mathbf{x}, \mathbf{y}) = \{(x_i,y_i);\hspace{1mm} i\hspace{-1mm} \in\hspace{-1mm} \{1,2,\ldots,N_b\}\}$ is the coordinate of locations of the $N_b$ BSs, and $F_i(r,\mathbf{x},\mathbf{y})$ and $f_i(r,\mathbf{x},\mathbf{y})$ are CDF and PDF of UE's distance from the BS in $i^{th}$ cell, respectively. Note that the $F_i(r,\mathbf{x},\mathbf{y})$ and $f_i(r,\mathbf{x},\mathbf{y})$ depend on the location of all BSs $(\mathbf{x},\mathbf{y})$ because the shape of a cell is determined using Voronoi tessellation as described in Section~\ref{sec:nwk_topo}. Using~\eqref{eq:cov} and \eqref{eq:PDF_multi}, the coverage probability of farthest UE from the BS in $i^{th}$ cell is expressed as:
\begin{align}
	P_{cov,i}^{far}=\int_r\exp\left(-\frac{\mathcal{T}\sigma^{2}r^\alpha}{P_t}\right)f_{far}^i\left(r,\mathbf{x},\mathbf{y}\right)\mathrm{d}r
\end{align}

\subsection{Optimization Formulation}\label{sec:opt_form}
In this chapter, our aim is to minimize the total power dissipation over the network while satisfying the given coverage constraint. The corresponding optimization problem (P0) can be formulated as: 
\begin{equation}\label{eq4}\nonumber
\begin{aligned}
\hspace{-2mm}(\text{P0})\hspace{-1mm}:\hspace{1mm}& \underset{N_b, P_t, \mathbf{x},\mathbf{y}}{\text{minimize}}
&&\hspace{-2mm}\mathrm N_b [\upepsilon_1 P_t + \upepsilon_2] \\
&\hspace{-8mm} \text{subject to}
&&\hspace{-8mm}C1\hspace{-0.5mm}: P_{\text{cov}, i}^{far}\geq 1-\epsilon, \; \forall i   \in  \{1, 2, \ldots ,N_b\},\\
&
&&\hspace{-8mm}C2\hspace{-0.5mm}: 0\leq N_b \leq N_{\text{max}},\\
&
&&\hspace{-8mm} C3\hspace{-0.5mm}: 0 \leq P_t \leq P_\text{max},\\
&
&&\hspace{-8mm}C4\hspace{-0.5mm}: 0 \leq  x_i \leq x_{\text{max}}, \; \forall i   \in  \{1, 2, \ldots ,N_b\},\\
&
&&\hspace{-8mm}C5\hspace{-0.5mm}:  0 \leq  y_i \leq y_{\text{max}}, \; \forall i  \in  \{1, 2, \ldots ,N_b\}, 
\end{aligned}
\end{equation} 
where $1-\epsilon$ (for $0\leq \epsilon \leq 1$) is the threshold coverage probability of the farthest UE that is satisfied under the constraint $C1$. $C2-C5$ are convex constraints that represent the bounds on $N_b$, $P_t$, $\mathbf{x}$, and $\mathbf{y}$, respectively. As the objective function of (P0) has integer variable $N_b$, the problem is nonconvex in the underlying variables. To obtain its optimal solution, we split the combinatorial problem into individual optimization of underlying variables $P_t$, $(\mathbf{x},\mathbf{y})$, and $N_b$ in following discussions. 

\subsection{Optimization of Transmit Power}
The constraint $C1$ is not tractable in the present form. To investigate, we simplify it using a tight approximation as follows. For satisfying the constraint above $90\%$ $\left(P_{\text{cov}, i}^{far}\geq 0.9,\; \forall i\right)$, the argument $\frac{\mathcal{T}\sigma^2 r^\alpha}{P_t}$ of the exponential term should be less than $0.1$ for a given PDF $f_{far}^i\left(r,\mathbf{x},\mathbf{y}\right)$. Therefore, $\exp\left(-\frac{\mathcal{T}\sigma^{2}r^\alpha}{P_t}\right)\approx 1-\frac{\mathcal{T}\sigma^{2}r^\alpha}{P_t}$ for $\frac{\mathcal{T}\sigma^{2}r^\alpha}{P_t}\leq 0.1$ with percentage error less than $0.05\%$. After applying this approximation in the constraint $C1$, we get
\begin{align}\label{eq:low_pow}
	P_t \geq \frac{\mathcal{T}\sigma^2}{\epsilon}\int_rr^\alpha f_{far}^i(r,\mathbf{x}, \mathbf{y})\mathrm{d}r 
\end{align} 
Though,~\eqref{eq:low_pow} gives the lower bound for the transmit power $P_t$ using the coverage probability in the $i^{th}$ cell, the optimal transmit power $P_t^*$ over the homogeneous network can be obtained by taking the maximum of the lower bound computed over different cells. The optimal transmit power $P_t^*$ can be mathematically expressed as:
\begin{align}\label{eq:opt_pow}
	P_t^* = \underset{i}{\max}\left\{\frac{\mathcal{T}\sigma^2}{\epsilon}\int_rr^\alpha f_{far}^i(r,\mathbf{x}, \mathbf{y})\mathrm{d}r\right\};\;\text{for}\; i  \in  \{1, 2, \ldots ,N_b\} 
\end{align}
After substituting $P_t^*$ into (P0), the problem (P1) can be formulated as:
\begin{equation}\nonumber
\begin{aligned}
\hspace{-3mm}(\text{P1})\hspace{-1mm}:\hspace{1mm}& \underset{N_b, \mathbf{x},\mathbf{y}}{\text{minimize}}
&&\hspace{-3mm}\mathrm N_b\left[c_b\cdot\underset{i}{\max}\left\{\int_rr^\alpha f_{far}^i(r,\mathbf{x}, \mathbf{y})\mathrm{d}r\right\}\hspace{-0.5mm}+\hspace{-0.5mm}\upepsilon_2\right] \\
&\hspace{-0mm} \text{subject to}
&&\hspace{-2mm} C2, C4, C5,
\end{aligned}
\end{equation}
where $c_b=\frac{\mathcal{T}\sigma^2 \upepsilon_1}{\epsilon}$. Next, using (P1), we optimize the location of the BSs for their given number $N_b$ over the specified field.

\section{Deployment Strategy for a Single BS}
To analyze the deployment of a single BS over the field, the optimization problem (P1) is equivalently expressed as:
\begin{equation}\label{eq6}\nonumber
\begin{aligned}
\hspace{-3mm}(\text{P2})\hspace{-1mm}:\hspace{1mm}& \underset{x,y}{\text{minimize}}
&&\hspace{-3mm}\int_rr^\alpha f_{far}(r,x,y)\mathrm{d}r \\
&\hspace{-0mm} \text{subject to}
&&\hspace{-2mm} \widehat{C4}:\;0\leq x\leq x_{\max},\; \widehat{C5}:\;0\leq y\leq y_{\max}
\end{aligned}
\end{equation}
Using (P2), we now describe the optimal deployment strategy of a single BS when the number of UEs are (i) large or (ii) moderate over the field.

\begin{figure}[!t]
	\sidecaption
	\includegraphics[scale=0.15]{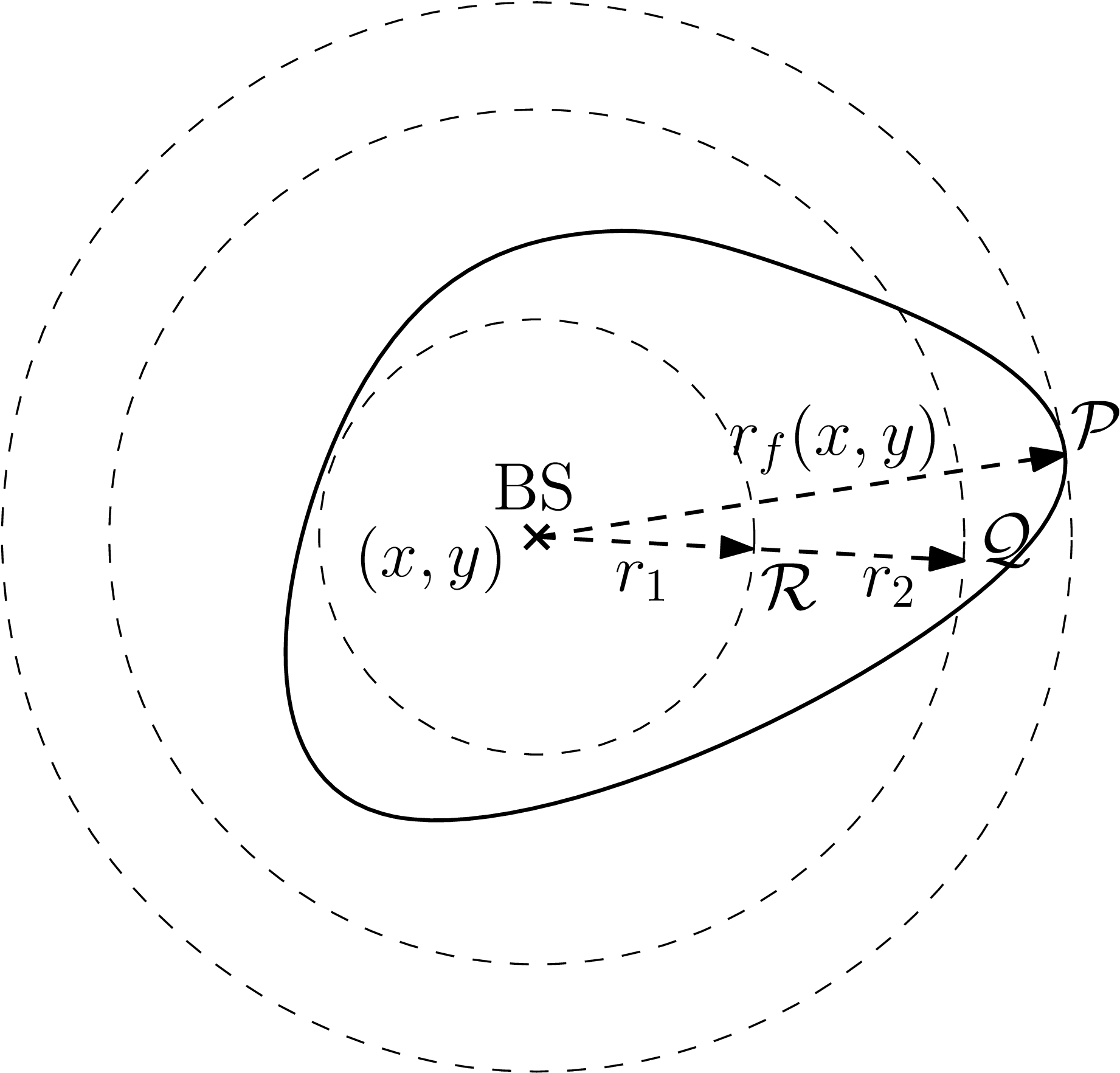}
	\caption{\small Farthest Euclidean Distance $r_f(x,y)$ from the BS located at a point $(x,y)$.}
	\label{fig:field_far}
\end{figure}

\subsection{For Large Number of UEs}
In the scenario, when the number of UEs is large, i.e., $N_u\rightarrow \infty$, the optimal location of the BS can be determined using Lemma~\ref{lemma1} as follows.
\begin{svgraybox}
\begin{lemma}\label{lemma1}
 When $N_u\rightarrow \infty$, then the PDF of distance of farthest UE from the BS is given as:
		\begin{align}\label{eq:pdf_del}
			\lim_{N_u\to\infty}f_{far}(r,x,y)=\delta(r-r_f(x,y)),
		\end{align}
		where $r_f(x,y)$ is the farthest Euclidean distance from the BS located at $(x,y)$ and $\delta(\cdot)$ is a Dirac delta function.
\end{lemma} 
\end{svgraybox}
\begin{proof}
	If we consider Fig.~\ref{fig:field_far}, the probability of lying of farthest UE near the point $\mathcal{P}$ can be computed using the distribution given by~\eqref{eq:dist_single}. Probability of lying of farthest UE near the point $\mathcal{P}$ at $r_f$ from the BS for $N_u\rightarrow \infty$ is given as:
	\begin{align}\label{eq:far_prob}\nonumber
		\lim_{\underset{N_u\to\infty}{\Delta\rightarrow 0^+}}& \mathrm{Pr}(r_f-\Delta<r \leq r_f)=\lim_{ \underset{N_u\to\infty}{\Delta\rightarrow 0^+}}[F_{\text{far}}(r_f,x,y)-F_{\text{far}}(r_f-\Delta, x,y)]\\
		&=\lim_{N_u\to\infty}  [F(r_f,x,y)]^{N_u}-\lim_{ \underset{N_u\to\infty}{\Delta\rightarrow 0^+}} [F(r_f-\Delta, x,y)]^{N_u}=1-0=1
	\end{align}
	Here note that $F(r,x,y)=1$ and $<1$ for $r=r_f$ and $<r_f$, respectively. Therefore, in~\eqref{eq:far_prob}, $\lim_{N_u\to\infty}  [F(r_f,x,y)]^{N_u}=1$ and $\lim_{ \underset{N_u\to\infty}{\Delta\rightarrow 0^+}} [F(r_f-\Delta, x,y)]^{N_u}=0$. Now we find the  probability of lying of farthest UE between the intermediate points $\mathcal{Q}$ at $r_2$ and $\mathcal{R}$ at $r_1$ as shown in Fig.~\ref{fig:field_far}. It can be expressed as:
	\begin{align}\label{eq:far_int}
		\lim_{N_u\to\infty} \mathrm{Pr}(r_1<r \leq r_2)=\lim_{N_u\to\infty}  [F(r_2,x,y)]^{N_u}-\lim_{N_u\to\infty} [F(r_1, x,y)]^{N_u}=0.
	\end{align}
	Therefore, from~\eqref{eq:far_prob} and \eqref{eq:far_int}, for $N_u\rightarrow \infty$, the farthest UE always lies at the farthest Euclidean distance from the BS. Hence from~\eqref{eq:far_prob}, $\lim_{N_u\to\infty}\int_{r_f-\Delta}^{r_f}f_{far}(r,x,y)\mathrm{d}r=\int_{r_f-\Delta}^{r_f}\delta(r-r_f)\mathrm{d}r$ that gives the PDF as expressed in~\eqref{eq:pdf_del}.
\end{proof}
For $N_u \rightarrow \infty$, if we substitute the obtained PDF in~\eqref{eq:pdf_del} into the objective function of (P2), the problem equivalently changed to the minimization of $\{r_f(x,y)\}^\alpha$ or $r_f(x,y)$ under the constraints $\widehat{C4}$ and $\widehat{C5}$. \textit{Based on it, it can be easily shown that the optimal location of a BS in a circular or in a regular polygon is at the center from which the farthest euclidean distance is minimum.}

\subsection{For Moderate Number of UEs}
\begin{figure}[!t]
	\sidecaption
	\includegraphics[scale=0.2]{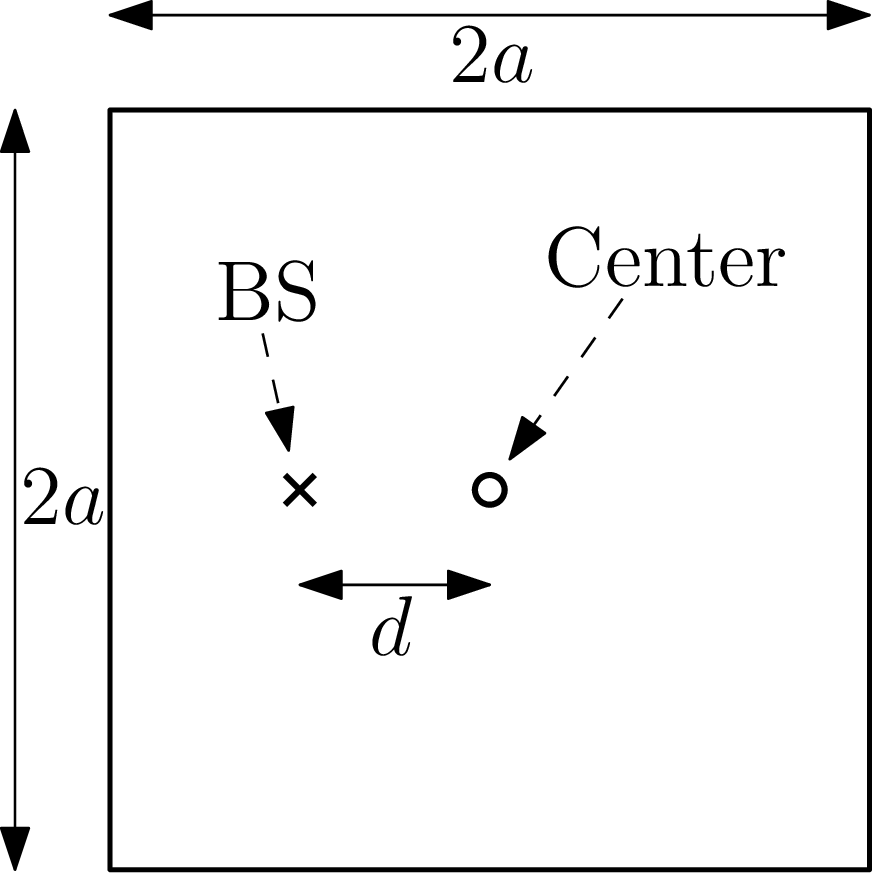}
	\caption{\small Optimal deployment strategy of a single BS in a square field when number of distributed UEs is moderate.}
	\label{fig:sq_field}
\end{figure}
\begin{figure}[!t]
	\sidecaption
	\includegraphics[scale=0.35]{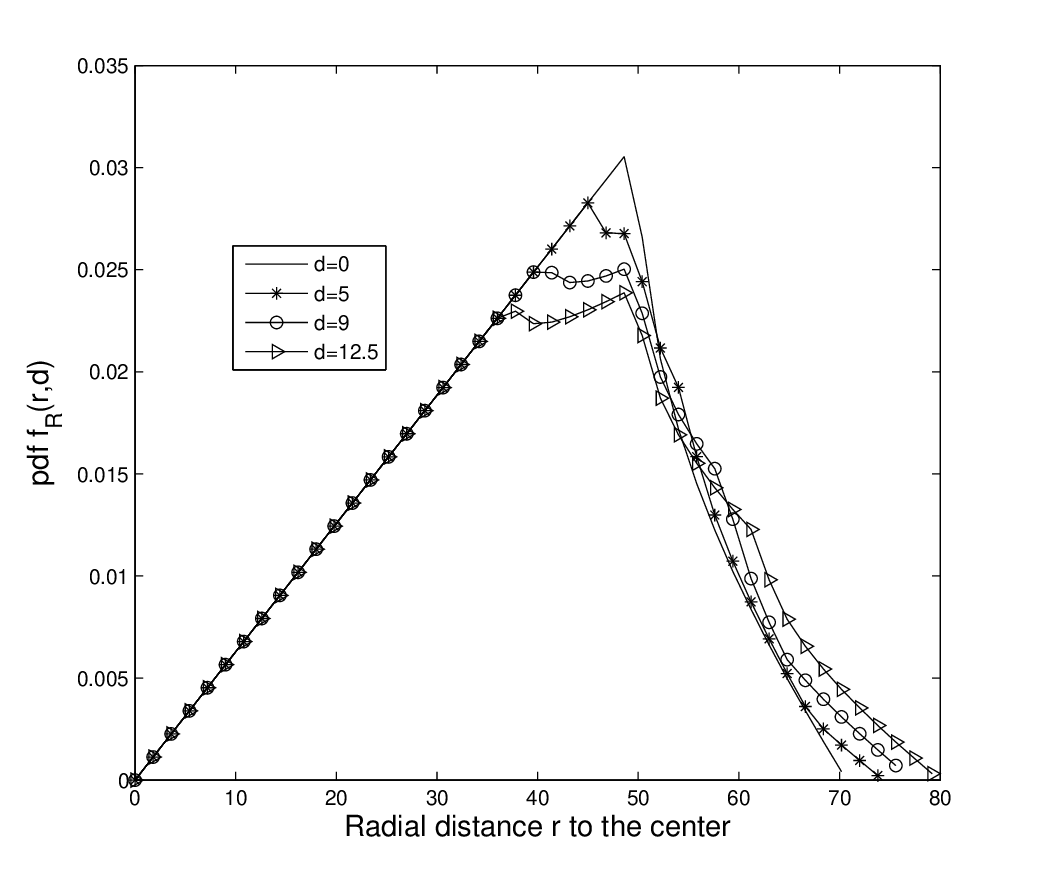}
	\caption{\small PDF $f(r,d)$ of distance $r$ of an UE from the BS located at distance $d$ from the center of the square field (cf. Fig.~\ref{fig:sq_field}).}
	\label{fig:sq_pdf}
\end{figure}
To obtain the optimal solution of (P2) for a moderate number of UEs present over the field, first, we investigate it for a square field as shown in Fig.~\ref{fig:sq_field}. Here, a BS is located at distance $d$ (leftward) from the center of the square field. The expression of PDF $f(r,d)$ of distance $r$ of a UE from the BS is derived in the appendix at the end of the chapter. The variation of $f(r,d)$ with $r$ for different location $d$ of the BS is numerically plotted in Fig.~\ref{fig:sq_pdf}. It shows that the peak of the PDF is highest when BS is located at $d=0$ (center of the square field) and it gradually decreases with $d$ and in contrary, the farthest Euclidean distance $r_f$ from the BS increases with $d$. From~\eqref{eq:CDF_PDF_Far}, PDF $f_{far}(r,d)$ of farthest UE distance also has the same variation and based on it, we find the optimal deployment strategy of a single BS in the square field using Lemma~\ref{lemma2} when $N_u$ is moderate.
\begin{figure}[!t]
	\sidecaption
	\includegraphics[scale=0.3]{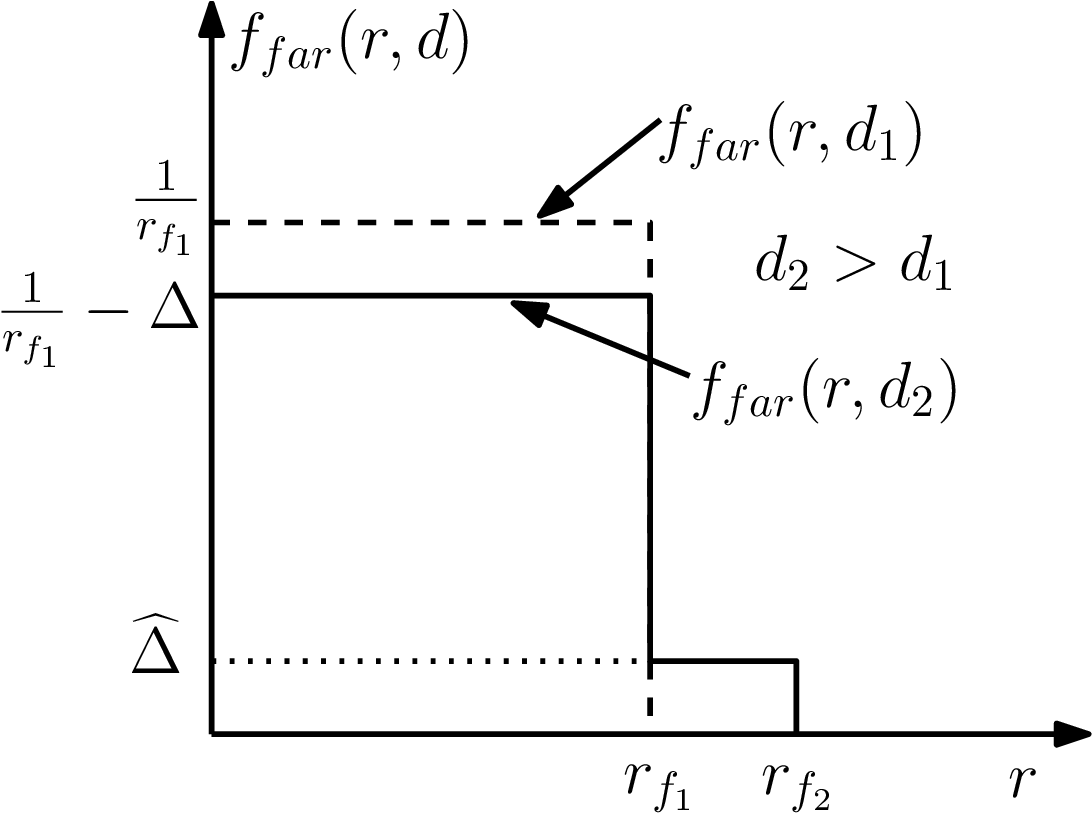}
	\caption{\small Diagram for describing the variation of peak and farthest Euclidean distance of $f(r,d)$ (cf. Fig.~\ref{fig:sq_pdf}) for $d=d_1$ and $d_2$, where $d_1<d_2$.}
	\label{fig:pdf_far_model}
\end{figure}
\begin{svgraybox}
\begin{lemma}\label{lemma2}
	If peak of PDF $f_{far}(r,d_1)$ is greater than the peak of $f_{far}(r,d_2)$, whereas the corresponding farthest Euclidean distance $r_{f_1}<r_{f_2}$ for $d_1<d_2$ as shown in Fig.~\ref{fig:pdf_far_model}, then the $N^{th}$ moment of the farthest UE distance under the two distribution satisfies:
	\begin{align}\label{eq:lemma2}
		\int_{0}^{r_{f_1}}r^N f_{far}(r,d_1)\mathrm{d}r < \int_{0}^{r_{f_2}}r^N f_{far}(r,d_2)\mathrm{d}r.
	\end{align} 
\end{lemma}  
\end{svgraybox}
\begin{proof}
	From Fig.~\ref{fig:pdf_far_model}, $\Big(\frac{1}{r_{f_1}}-\Delta\Big)r_{f_1}+\widehat{\Delta} (r_{f_2}-r_{f_1})=1$ that gives $\widehat{\Delta}=\frac{\Delta\hspace{0.5mm} r_{f_1}}{r_{f_2}-r_{f_1}}$ for $d_1 < d_2$. Using it, the result in~\eqref{eq:lemma2} can be proved as:
	\begin{align}\label{eq:lemma2_proof}\nonumber
		\int_{0}^{r_{f_2}}&r^N f_{far}(r,d_2)\mathrm{d}r=\int_{0}^{r_{f_1}}r^N\left(\frac{1}{r_{f_1}}-1\right)\mathrm{d}r+\int_{r_{f_1}}^{r_{f_2}}r^N \widehat{\Delta}\mathrm{d}r\\\nonumber&
		=\left(\frac{1}{r_{f_1}}-\Delta\right)\frac{r_{f_1}^{N+1}}{N+1}+\frac{\Delta\hspace{0.5mm} r_{f_1}}{r_{f_2}-r_{f_1}}\frac{r_{f_2}^{N+1}-r_{f_1}^{N+1}}{N+1}\\\nonumber&
		=\int_{0}^{r_{f_1}}r^N f_{far}(r,d_1)\mathrm{d}r+\frac{\Delta r_{f_1}}{N+1}(r_{f_2}^N+r_{f_2}^{N-1}r_{f_1}+\cdots+r_{f_2}r_{f_1}^{N-1})\\&
		>\int_{0}^{r_{f_1}}r^N f_{far}(r,d_1)\mathrm{d}r
	\end{align}
	Therefore, the $N^{th}$ moment has the minimum value at $d=0$ (center) of the square field. 
\end{proof} 
Thus, from~\eqref{eq:lemma2_proof}, the objective function of (P2) achieves the minimum value when the BS is located at the center of the square field. \textit{Similarly, it can be shown that for moderate $N_u$, the center of any regular polygon or circular field is the optimal location for the deployment of a single BS.} Besides, at the optimal location, the $N^{th}$ moment can be reduced by minimizing the farthest Euclidean distance $r_f$ of the fields the same as the case of large $N_u$.

\section{Deployment Strategy for Multiple BSs}
\begin{figure}[!t]
	\sidecaption
	\includegraphics[scale=0.18]{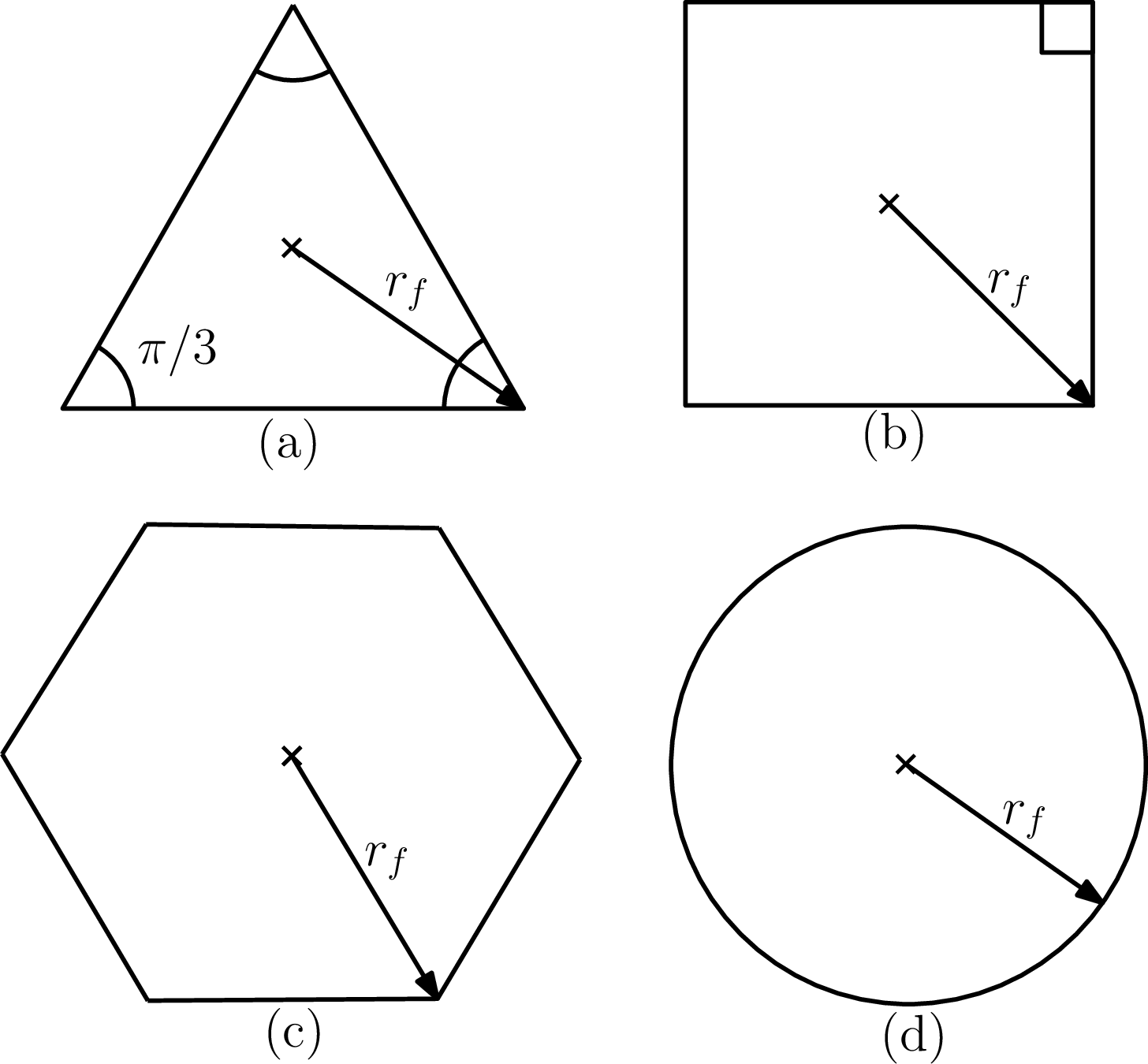}
	\caption{\small Comparison of different shape of the cells having same farthest Euclidean distance with respect to coverage and coverage hole.}
	\label{fig:cell_shape}
\end{figure}
In order to deploy multiple BSs over the square and circular field, we need to find the optimum shape of the generated cells with respect to coverage of UEs and coverage holes. To compare the different shapes of the cells, we assume that the UEs are distributed using HPPP over a large field. Note that although our analysis is based on BPP, it converges into HPPP when the size of the field becomes large. If the cells as shown in Fig.~\ref{fig:cell_shape} have same farthest Euclidean distance $r_f$ from their associated BSs, they can be compared in coverage using Lemma~\ref{lemma3}.
\begin{svgraybox}
\begin{lemma}\label{lemma3}
	If the number of UEs over a field is deployed using HPPP with density $\lambda$, then the coverage to average number of UEs in square, hexagonal, and circular cells is $53.96\%$, $100\%$, and $141.84\%$ more than triangular cell of same farthest Euclidean distance. 
\end{lemma}
\end{svgraybox}
\begin{proof}
	As the area of a equilateral triangular cell with the farthest Euclidean distance $r_f$ is $\frac{3\sqrt{3}r_f^2}{4}$, average number of UEs $N_{u,\mathcal{T}}=\frac{3\sqrt{3}r_f^2}{4}\lambda$. Likewise, the average number of UEs $N_{u,\mathcal{S}}$, $N_{u,\mathcal{H}}$, and $N_{u,\mathcal{C}}$ in square, hexagonal, and circular cells are $2r_f^2\lambda$,  $\frac{3\sqrt{3}r_f^2}{2}\lambda$, and $\pi r_f^2 \lambda$, respectively. Percentage coverage improvement in square field compared to triangular cell is $\Delta_{u,\mathcal{T}\to \mathcal{S}}=\frac{N_{u,\mathcal{S}}-N_{u,\mathcal{T}}}{N_{u,\mathcal{T}}}\times 100=53.96\%$. Similarly, the improvement in hexagonal and circular cells against the triangular is $\Delta_{u,\mathcal{T}\to \mathcal{H}}=100\%$ and $\Delta_{u,\mathcal{T}\to \mathcal{C}}=141.84\%$, respectively.  	
\end{proof}
From Lemma~\ref{lemma3}, the circular cell has the best coverage capability, but it creates the coverage holes over the field~\cite{rap11}. Therefore, hexagonal cell better in coverage of UEs as well as in coverage hole performances in a large field. However, in a finite square or circular field, hexagonal cells create holes at the boundaries, therefore we take square or rectangular cells in a square field and arc or triangular cells in a circular field to avoid coverage holes as shown in Figs.~\ref{fig:square_field_cells} and \ref{fig:circle_field_cells}. Now, the optimal deployment of multiple BSs over the fields can be determined using the optimization problem (P1) when $N_u$ is large or moderate.

\subsection{For large $N_u$}\label{sec:multi_large_Nu}
\begin{figure}[!t]
	\sidecaption
	\includegraphics[scale=0.45]{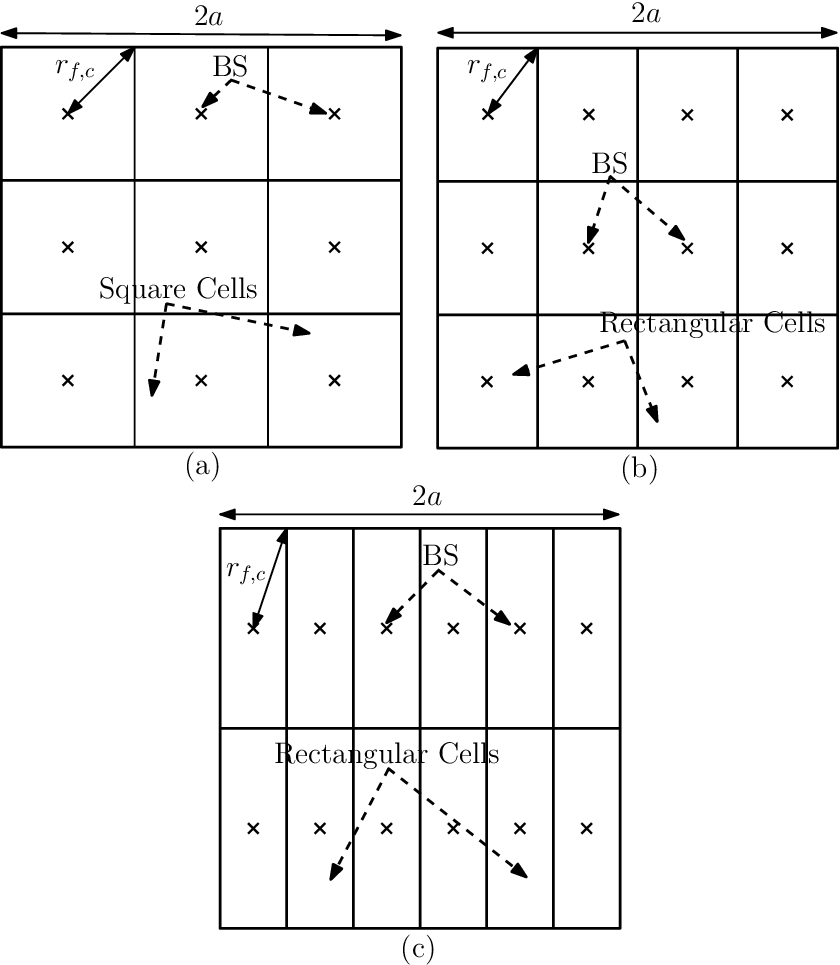}
	\caption{\small Square or rectangular cells in a square field due to different location of the BSs. (a) $N_b=9$, $m=3$, and $n=3$. (b) $N_b=12$, $m=3$, and $n=4$. (c) $N_b=12$, $m=2$, and $n=6$.}
	\label{fig:square_field_cells}
\end{figure}

For a large number of UEs $(N_u\rightarrow \infty)$ in a finite field, the number of UEs in each cell of the field is also large. Therefore, from Lemma~\ref{lemma1}, for a large number of UEs in $i^{th}$ cell $(N_{u,i}\rightarrow \infty)$, the corresponding PDF of the distance of farthest UE from the associated BS can be expressed as:
\begin{align}\label{eq:pdf_large_multi}
	f_{far}^i(r,\mathbf{x},\mathbf{y})=\delta(r-r_{f,i}(\mathbf{x},\mathbf{y})),
\end{align}
where $r_{f,i}$ is farthest Euclidean in the $i^{th}$ cell which is function of location $(\mathbf{x},\mathbf{y})$ of all BSs in the field. After substituting~\eqref{eq:pdf_large_multi} into the objective function of (P1), we get the optimization problem:
\begin{equation}\nonumber
\begin{aligned}
\hspace{-3mm}(\text{P3})\hspace{-1mm}:\hspace{1mm}& \underset{N_b, \mathbf{x},\mathbf{y}}{\text{minimize}}
&&\hspace{-3mm}\mathrm N_b\left[c_b\cdot\underset{i}{\max}\left\{[r_{u,i}(\mathbf{x},\mathbf{y})]^\alpha\right\}\hspace{-0.5mm}+\hspace{-0.5mm}\upepsilon_2\right] \\
&\hspace{-0mm} \text{subject to}
&&\hspace{-2mm} C2, C4, C5,
\end{aligned}
\end{equation}
Now, using (P3) we find the optimal deployment of $N_b$ BSs over the square field. For a given $N_b$, BSs' locations can be optimized by minimizing $\underset{i}{\max}\left\{r_{f,i}(\mathbf{x},\mathbf{y})\right\}$ under the constraints $C4$ and $C5$. As described above, the optimal shape of the cells with respect to coverage of UEs and coverage hole is square or rectangular with equal farthest Euclidean distance $(r_{f,i}=r_{f,c}\;\forall i)$ from their BSs optimally located at the center of the cells as shown in Fig.~\ref{fig:square_field_cells}. 
As a result, minimization of $\underset{i}{\max}\left\{r_{f,i}(\mathbf{x},\mathbf{y})\right\}$ is reduced to minimizing $r_{f,c}$ using a different arrangement of the cells along the length and the width of the field. The optimal arrangement can be determined using Lemma~\ref{lemma4} as follows.
\begin{svgraybox}
\begin{lemma}\label{lemma4}
	For $N_b=m\times n$ number of BSs are deployed in their respective cells over a square field, the minimum value of farthest Euclidean distance $r_{f,c}$ is achieved at $m=n=\sqrt{N_b}$ or minimum $|m-n|$ when $\sqrt{N_b}$ is an integer or not an integer, respectively.
\end{lemma}
\end{svgraybox}
\begin{proof}
	If we relax the integer value of number of rows $m$ and number of columns $n$, then we assume that $n=m-\omega$, where $\omega\geq 0$. The farthest Euclidean distance $r_{f,c}=\sqrt{\Big(\frac{a}{m}\Big)^2+\Big(\frac{a}{m-\omega}\Big)^2}$, where $2a$ is the side length of the square field as in Fig.~\ref{fig:square_field_cells}. Here, $r_{f,c}$ is equivalently minimized by minimizing $\mathcal{D}\triangleq\frac{1}{m^2}+\frac{1}{(m-\omega)^2}$. As $N_b=m(m-\omega)$, we get $m=\frac{\omega+\sqrt{\omega^2+4N_B}}{2}$. After substitution of $m$, $\mathcal{D}$ can be further expressed as $\mathcal{D}=\frac{\omega^2+2N_B}{N_B^2}$. As, $\frac{\partial \mathcal{D}}{\partial \omega}=\frac{2\omega}{N_B^2}$ and $\frac{\partial^2 \mathcal{D}}{\partial \omega^2}=\frac{2}{N_B^2}>0$, $\mathcal{D}$ is convex. Thus, $\mathcal{D}$ and $r_{f,c}$ achieves its minimum value at $\omega=m-n=0$ that gives $m=n=\sqrt{N_b}$ when $\sqrt{N_b}$ is an integer. Otherwise, factorize $N_b$ into $m$ and $n$ such that $|\omega|=|m-n|$ is minimum.  
\end{proof}
For example, in Figs.~\ref{fig:square_field_cells}(b) and \ref{fig:square_field_cells}(c), through the total number of BSs $N_b=12$, but in Fig.~\ref{fig:square_field_cells}(b), the number of rows denoted as $m$ is $3$ and number of columns denoted as $n$ is $4$. Whereas in Fig.~\ref{fig:square_field_cells}(c), $m=2$ and $n=6$. As $\sqrt{N_b}=\sqrt{12}=3.46$ is not an integer, we factorize $N_b=12$ into $m$ and $n$ such that the difference $|m-n|$ is minimum (cf. Lemma~\ref{lemma4}). The arrangement in Fig.~\ref{fig:square_field_cells}(b) is optimum because $|m-n|=|3-4|=1$ is minimum possible value as compared to the arrangement in Fig.~\ref{fig:square_field_cells}(c) where $|m-n|=|2-6|=4$.

\begin{figure}[!t]
	\sidecaption
	\includegraphics[scale=0.2]{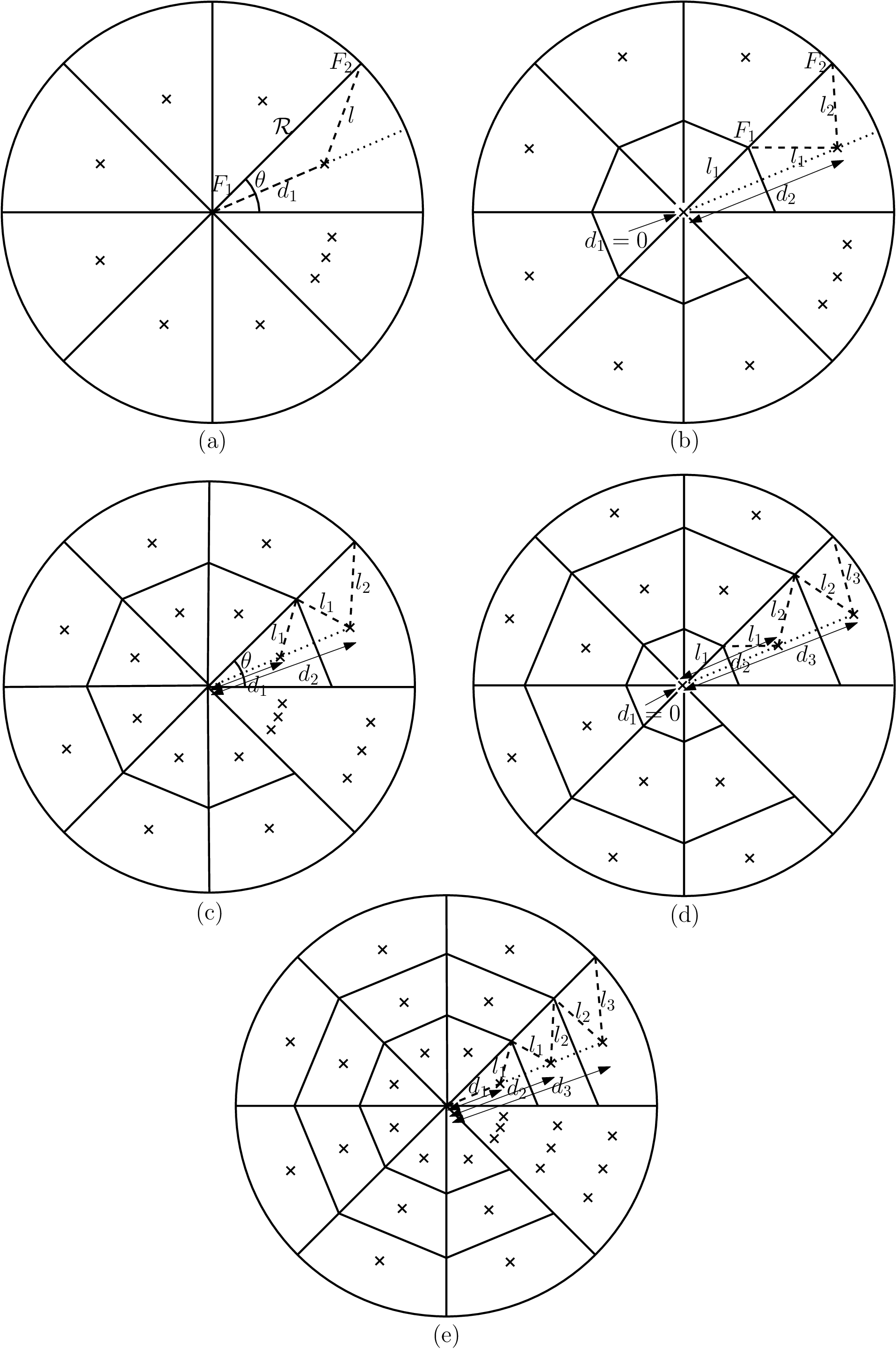}
	\caption{\small Arc, rhombus, and triangular cells in a circular field due to different location of the BSs. (a) $N_b=q$, (b) $N_b=q+1$, (c) $N_b=2q$, (d) $N_b=2q+1$, and (e) $N_b=3q$.}
	\label{fig:circle_field_cells}
\end{figure}

In case of circular field, the BSs are deployed in radial and angular directions as shown in Fig.~\ref{fig:circle_field_cells}. Based on it, we define the arrangement as $pq+t$, where $t=0$ and $t=1$ are for absence  and presence of a BS at the center of the field, $q$ is the number of arcs in angular direction, and $p$ is the number of cells in each arc along the radial direction. The deployment of BSs in each arc is same, therefore, we study the optimal deployment of BSs in an arc of the field. Note that in each arc, the BSs are located at the symmetric line as shown in Fig.~\ref{fig:circle_field_cells}. As $N_u\rightarrow \infty$, the optimal deployment is determined using farthest Euclidean distance of the cells. For a given $N_b$, the optimal deployment can be determined by $\min\;\underset{i}{\max}\{r_{f,i}\}$ for $i\in\{1-t,2-t,\cdots,p\}$. As described in~\cite[Section IV]{pra13}, using the trigonometric relationship in the particular arc of Figs.~\ref{fig:circle_field_cells}(a), \ref{fig:circle_field_cells}(b), \ref{fig:circle_field_cells}(c), \ref{fig:circle_field_cells}(d), and \ref{fig:circle_field_cells}(e), we can find the optimal arrangement (type), $\min\;\underset{i}{\max}\{r_{f,i}\}$, optimal location of BSs for a given $N_b$ which are listed in Table~\ref{tab:table1} and Table~\ref{tab:table2}. 

\begin{table*}[!t]
	\def\arraystretch{0.86}
	\centering
	\vspace{-3mm}\caption{For a given number of BSs $N_b$, optimal arrangement (type) of BSs and corresponding optimal location in an arc of the circular field.}
	\label{tab:table1}
	\setlength\tabcolsep{3pt}
	\scalebox{1.105}
	{\begin{tabular}{|c|c|c|}
			\hline
			Given $N_b$ & Optimal & Optimum location of the BSs in an arc    \\
			 &Type&\\
			\hline 
			$N_b=3$ &$q$  & $d_1^{*}=\mathcal{R}\cos\left(\frac{\pi}{N_b}\right)$\\ \cline{1-1}\cline{3-3}
			$N_b \in \{4,5,6\}$ && $d_1^{*}=\frac{\mathcal{R}}{2\cos\left(\frac{\pi}{N_b}\right)}$ \\
			\hline  
			$N_b \in \{7,8,\ldots,17\}\cup \{19\}$ & $q+1$ & $d_0^{*}=0, d_1^{*}= \frac{2\mathcal{R}\cos\left(\frac{\pi}{N_b-1}\right)}{4\cos^2\left(\frac{\pi}{N_b-1}\right)-1}$\\
			\hline
			$ N_b\in\{18, 20,\ldots,44\}$ & $2q$ & $d_1^{*}=\frac{\mathcal{R}}{4\cos\left(\frac{2\pi}{N_b}\right)\cos\left(\frac{4\pi}{N_b}\right)}$,\\
			&&$d_2^{*}=\frac{\mathcal{R}\left(1+\cos\left(\frac{4\pi}{N_b}\right)\right)}{4\cos\left(\frac{2\pi}{N_b}\right)\cos\left(\frac{4\pi}{N_b}\right)}$\\    
			\hline
			$N_b \in \{21,23,\ldots,45\}$ & $2q+1$ & $d_0^{*}=0$, $d_1^{*}=\frac{2\mathcal{R}\left(1+2\cos\left(\frac{4\pi}{N_b-1}\right)\right)\cos\left(\frac{2\pi}{N_b-1}\right)}{16\cos^2\left(\frac{2\pi}{N_b-1}\right)\cos^2\left(\frac{4\pi}{N_b-1}\right)-1}$,\\ && $d_2^{*}=\frac{4\mathcal{R}\left(1+2\cos\left(\frac{4\pi}{N_b-1}\right)\right)\cos\left(\frac{4\pi}{N_b-1}\right)\cos\left(\frac{2\pi}{N_b-1}\right)}{16\cos^2\left(\frac{2\pi}{N_b-1}\right)\cos^2\left(\frac{4\pi}{N_B-1}\right)-1}$\\
			\hline
			$N_b\in \{48,51,\ldots\}$ & $3q$ & $d_1^{*}=\frac{\mathcal{R}\cos\left(\frac{3\pi}{N_b}\right
				)}{\left(2\cos\left(\frac{6\pi}{N_b}\right)+1\right)\left(\cos\left(\frac{12\pi}{N_b}\right)+\cos\left(\frac{6\pi}{N_b}\right)\right)}$,\\&& $d_2^{*}=\frac{\mathcal{R}\cos\left(\frac{3\pi}{N_b}\right)\left(1+2\cos\left(\frac{6\pi}{N_b}\right)\right)}{\left(2\cos\left(\frac{6\pi}{N_b}\right)+1\right)\left(\cos\left(\frac{12\pi}{N_b}\right)+\cos\left(\frac{6\pi}{N_b}\right)\right)}$, \\ && $d_3^{*}=\frac{\mathcal{R}\cos\left(\frac{3\pi}{N_b}\right)\left(1+2\cos\left(\frac{6\pi}{N_b}\right)+2\cos\left(\frac{9\pi}{N_b}\right)\right)}{\left(2\cos\left(\frac{6\pi}{N_b}\right)+1\right)\left(\cos\left(\frac{12\pi}{N_b}\right)+\cos\left(\frac{6\pi}{N_b}\right)\right)}$\\
			\hline
	\end{tabular}}
\end{table*} 

\begin{table*}[!t]
	\def\arraystretch{0.86}
	\centering
	\vspace{-3mm}\caption{For a given $N_b$, optimal arrangement (type) of BSs and corresponding minimum of maximum of farthest Euclidean distance over the cells in an arc of the circular field.}
	\label{tab:table2}
	\setlength\tabcolsep{3pt}
	\scalebox{1.105}
	{\begin{tabular}{ |c|c|c| }
			\hline
			Given $N_b$ & Optimal   & Minimum of $\underset{i}{\max}\;\{r_{f,i}\}$ \\
			& Type &\\
			\hline 
			$N_b=3$ & $q$ & $\mathcal{R}\sin\left(\frac{\pi}{N_b}\right)$\\ \cline{1-1}\cline{3-3}
			$N_b \in \{4,5,6\}$ && $\frac{\mathcal{R}}{2\cos\left(\frac{\pi}{N_b}\right)}$ \\
			\hline  
			$N_b \in \{7,8,\ldots,17\}\cup \{19\}$ & $q+1$ &$ \frac{\mathcal{R}}{4\cos^2\left(\frac{\pi}{N_b-1}\right)-1}$\\
			\hline
			$ N_b\in\{18, 20,\ldots,44\}$ & $2q$ &$\frac{\mathcal{R}}{4\cos\left(\frac{2\pi}{N_b}\right)\cos(\frac{4\pi}{N_b})}$\\   
			\hline
			$N_b \in \{21,23,\ldots,45\}$ & $2q+1$ & $\frac{\mathcal{R}\left(1+2\cos\left(\frac{4\pi}{N_b-1}\right)\right)}{16\cos^2\left(\frac{2\pi}{N_b-1}\right)\cos^2\left(\frac{4\pi}{N_b-1}\right)-1}$\\
			\hline
			$N_b\in \{48,51,\ldots\}$ & $3q$ & $\frac{\mathcal{R}\cos\left(\frac{3\pi}{N_b}\right)}{\left(2\cos\left(\frac{6\pi}{N_b}\right)+1\right)\left(\cos\left(\frac{12\pi}{N_b}\right)+\cos\left(\frac{6\pi}{N_b}\right)\right)}$ \\
			\hline
	\end{tabular}}
\end{table*} 

\subsection{For Moderate $N_u$}
When $N_u$ is moderate, the number of UEs occurring in a cell depends of the occurrence of UEs in other cells. As described above, square or rectangular cells are optimal in the square field and the optimal location of the BSs is at the center of the cells even for moderate $N_u$. Therefore, for a given $N_b$, the optimal deployment of BSs is based on minimization of farthest Euclidean distance over cells the same as the case of large $N_u$. Also, it has been shown in~\cite{pra13} that deployment based on the minimization of the farthest euclidean distance over the cells is acceptable in the circular field within negligible root mean square error (RMSE). Thus, for $N_u \rightarrow \infty$, the optimal deployment strategy of given BSs  over the square and circular fields is the same as described in Section~\ref{sec:multi_large_Nu}.

\section{Joint Optimization}
In this section, we study about the joint optimization of number of BSs $N_b$ and their deployment (arrangement of cells) in both square and circular field.

\subsection{Optimization over the Square Field} 
To jointly optimize the number of BSs $N_b$ and their location over the square field for $N_u\rightarrow \infty$, (P3) can be simplified as:
\begin{equation}\nonumber
\begin{aligned}
\hspace{-3mm}(\text{P4})\hspace{-1mm}:\hspace{1mm}& \underset{N_b, m, n}{\text{minimize}}
&&\hspace{-3mm}\mathrm N_b\left[\widehat{c}_b\cdot\left(\frac{1}{m^2}+\frac{1}{n^2}\right)^{\frac{\alpha}{2}} \hspace{-0.5mm}+\hspace{-0.5mm}\upepsilon_2\right], \\
&\hspace{-0mm} \text{subject to}
&&\hspace{-2mm} C2,\; \overline{C4}:1\leq m \leq N_b,\; \overline{C5}: 1 \leq n \leq N_b,
\end{aligned}
\end{equation}
where $\widehat{c}_b=a^\alpha c_b$ and $2a$ is side length of the square field. Iteratively, the optimal value of $N_b$, $m$, and $n$ can be obtained as follows. We start from $N_b=1$ and for each $N_b\in\{1,2,\cdots, N_{\max}\}$, optimal value of $m$ and $n$ is calculated using Lemma~\ref{lemma4}, thereafter we find the corresponding total power consumption using the objective function of (P4). The obtained total power consumption for different $N_b$ is compared and that $N_b$ is set to optimal number of BSs $N_b^*$ which gives minimum value of total power consumption and corresponding $m$ and $n$ are set as optimal number of rows $m^*$ and optimal number of columns $n^*$, respectively.  

For moderate $N_u$, the joint optimization can by obtained by solving (P1) which can be expressed for the square field as:
\begin{equation}\nonumber
\begin{aligned}
\hspace{-3mm}(\text{P5})\hspace{-1mm}:\hspace{1mm}& \underset{N_b, m, n}{\text{minimize}}
&&\hspace{-3mm}\mathrm N_b\left[c_b\cdot\int_{0}^{\sqrt{\left(\frac{a}{m}\right)^2+\left(\frac{a}{n}\right)^2}}r^\alpha f_{far}^c(r,d=0)\mathrm{d}r\hspace{-0.5mm}+\hspace{-0.5mm}\upepsilon_2\right] \\
&\hspace{-0mm} \text{subject to}
&&\hspace{-2mm} C2, \overline{C4}, \overline{C5},
\end{aligned}
\end{equation}
where $d=0$ denotes that BS in each cell is located at the center and $f_{far}^c(r,d=0)$ is obtained using~\eqref{eq:PDF_multi}, \eqref{eq36}, and \eqref{eq37}. Here, superscript $c$ in the distribution denotes that the distribution in each cell is same because the cells are identical in the square field. Again, for the optimal solution, we start from $N_b=1$, and for each $N_b$, optimal value of $m$ and $n$ is obtained using Lemma~\ref{lemma4}. From it, the total power consumption is computed using the objective function of (P5) for each $N_b$. The $N_b$ which gives the minimum value of power consumption is set as $N_b^*$ and corresponding $m$ and $n$ are set as $m^*$ and $n^*$, respectively.

\subsection{Optimization over the Circular Field}
In case of joint optimization over the circular field, we find the optimal arrangement (type) and corresponding optimal location and $\min\;\underset{i}{\max}\{r_{f,i}\}$ using Tables~\ref{tab:table1} and~\ref{tab:table2} for each $N_b$. For $N_u\rightarrow \infty$ and moderate $N_u$, we find the total power consumption using the objective function of (P3)
and (P1) respectively. The $N_b$ which gives the minimum power consumption is set to $N_b^*$ and corresponding arrangement (type) and locations are set as the optimal value. 

\section{Numerical Results and Discussions}
In this section, we describe the obtained numerical results for the discoursed analysis where the default value of the used system parameters for the results are listed in Table~\ref{tab:table3}. Using it, we find the design insights on the transmit power $P_t$, number of BSs $N_b$, and total power consumption with variation of system parameters.

\begin{table}[!h]
	\caption{\small List of system parameters with their default values.}
	\label{tab:table3}
	\centering
	\begin{tabular}{lc}
		
		System Parameter & Value \\
		\hline
		Side length of the square field, $2a$ & 1000 m \\
		Radius of the circular field, $R$ & 500 m \\
		AWGN noise power, $\sigma^2$ & $-70$ dBm\\
		Mean of exponential distribution, $\mu$ & $1$ \\
		Threshold SNR, $\mathcal{T}$ & -10 dB \\
		Path loss exponent, $\alpha$ & $4$\\
		Scaling parameter, $\upepsilon_1$ & $5.5$ \\
		Electronics, processing, and battery\\ backup power losses, $\upepsilon_2$ & $32$ W \\
		Maximum transmit power, $P_{\max}$ & $5$ W \\
		Maximum number of BSs, $N_{\max}$ & 35 \\
	\end{tabular}
\end{table}

\begin{figure}[!t]
	\sidecaption
	\includegraphics[scale=0.6]{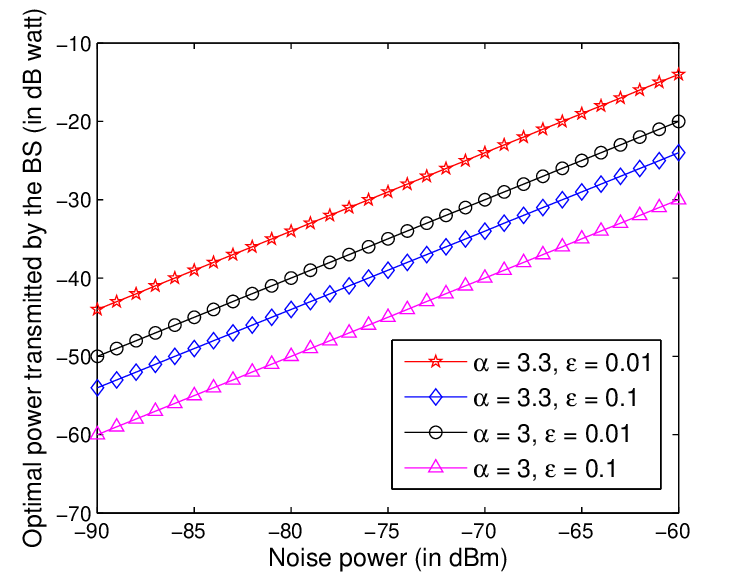}
	\caption{\small Optimal transmit power $P_t^*$ with noise power for different values of $\alpha$ and $\epsilon$ over the circular field.}
	\label{fig:opt_pow_circular}
\end{figure}

\begin{figure}[!t]
	\sidecaption
	\includegraphics[scale=0.59]{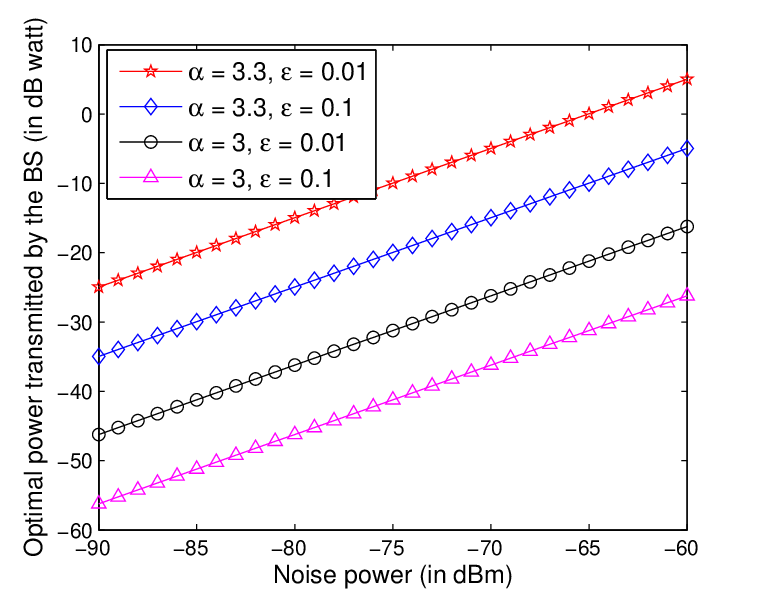}
	\caption{\small Optimal transmit power $P_t^*$ with noise power for different values of $\alpha$ and $\epsilon$ over the square field.}
	\label{fig:opt_pow_square}
\end{figure}

Figs.~\ref{fig:opt_pow_circular} and \ref{fig:opt_pow_square} describe the variation of optimal transmit power $P_t^*$ with noise power $\sigma^2$ for different values of path loss exponent $\alpha$ and acceptable tolerance $\epsilon$. It can be observed that to increase the coverage from $90\%$ to $99\%$, we require to increase $P_t^*$ on average by $10$ dB Watt over both the fields. On the other hand, if $\alpha$ is changed from $3$ to $3.3$, $P_t^*$ is increased by $6$ dB Watt over the circular field, whereas it is increased by $20$ dB Watt over the square field. Therefore, the circular field is more energy-efficient than the square field.  

\begin{figure}[!t]
	\sidecaption
	\includegraphics[scale=0.7]{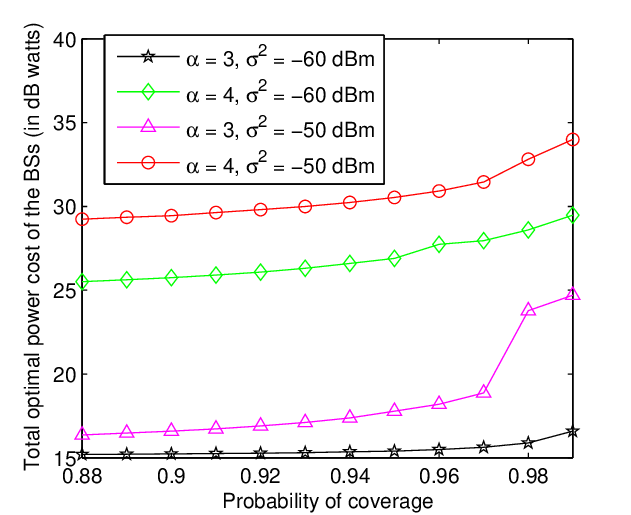}
	\caption{\small Minimized total power consumption with coverage probability for different value of $\alpha$ and $\sigma^2$ over the circular field.}
	\label{fig:total_pow_circular}
\end{figure}

\begin{figure}[!t]
	\sidecaption
	\includegraphics[scale=0.7]{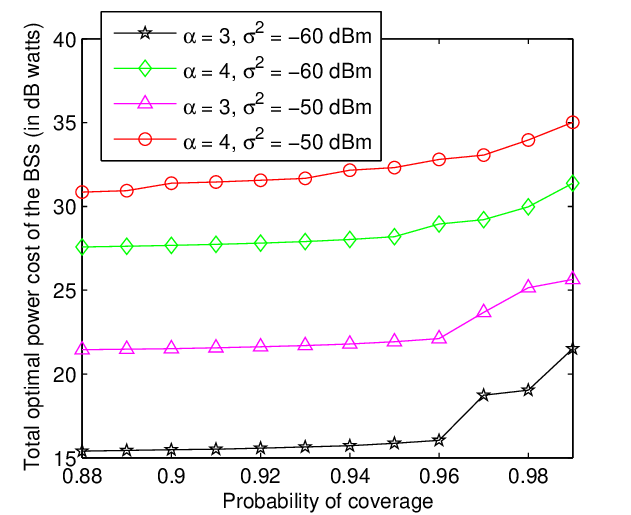}
	\caption{\small Minimized total power consumption with coverage probability for different value of $\alpha$ and $\sigma^2$ over the square field.}
	\label{fig:total_pow_square}
\end{figure}

In Figs.~\ref{fig:total_pow_circular} and \ref{fig:total_pow_square}, the obtained results depict the total power consumption with coverage probability over the circular and square field, respectively. Here, in the circular field, the increment in the total power consumption at $\alpha=4$ is more than $\alpha=3$ when $\sigma^2$ changes from $-60$ dB to $-50$ dB. In contrast, over the square field, higher increment takes place at $\alpha=3$ for the same change in $\sigma^2$. Therefore, the energy saving in circular field is highly sensitive with noise for higher path loss exponent, but in square field, the sensitivity is high for its low value. 

\begin{figure}[!t]
	\sidecaption
	\includegraphics[scale=0.7]{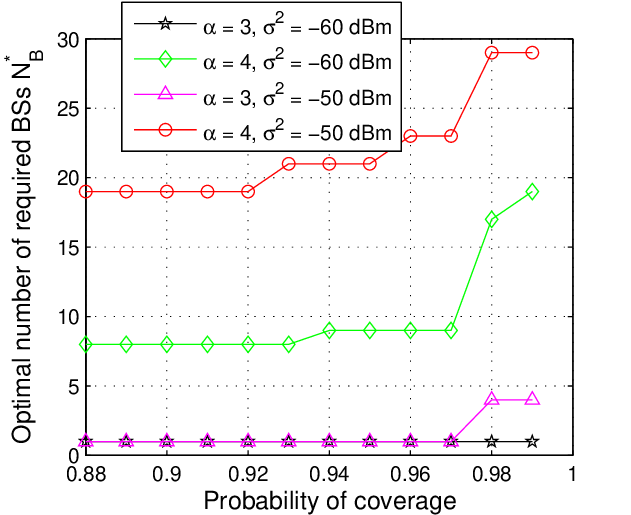}
	\caption{\small Optimal number of BSs $N_b^*$ with coverage probability for different value of $\alpha$ and $\sigma^2$ over the circular field.}
	\label{fig:optimal_Nb_circular}
\end{figure}

\begin{figure}[!t]
	\sidecaption
	\includegraphics[scale=0.7]{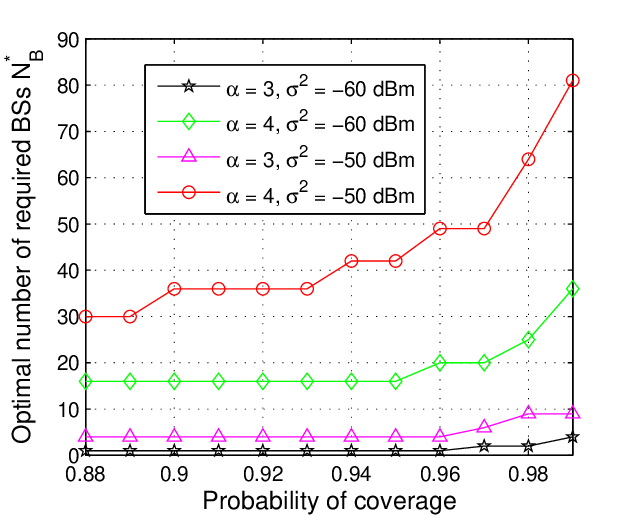}
	\caption{\small Optimal number of BSs $N_b^*$ with coverage probability for different value of $\alpha$ and $\sigma^2$ over the square field.}
	\label{fig:optimal_Nb_square}
\end{figure}

\begin{figure}[!t]
	\sidecaption
	\includegraphics[scale=0.69]{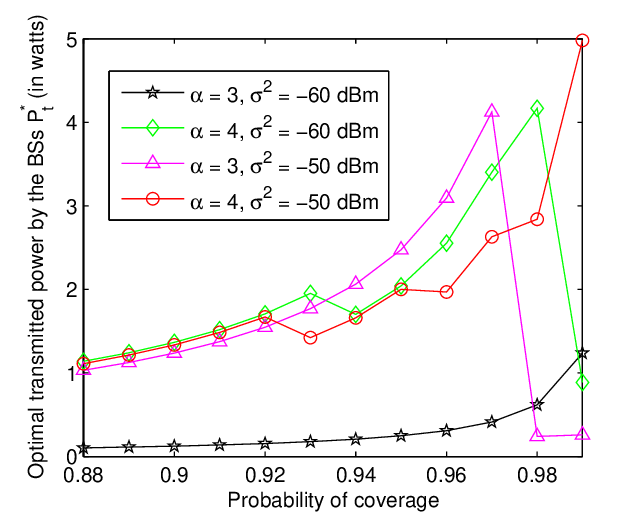}
	\caption{\small Optimal transmit power $P_t^*$ with coverage probability for different value of $\alpha$ and $\sigma^2$ over the circular field.}
	\label{fig:optimal_Pt_circular}
\end{figure}

\begin{figure}[!t]
	\sidecaption
	\includegraphics[scale=0.7]{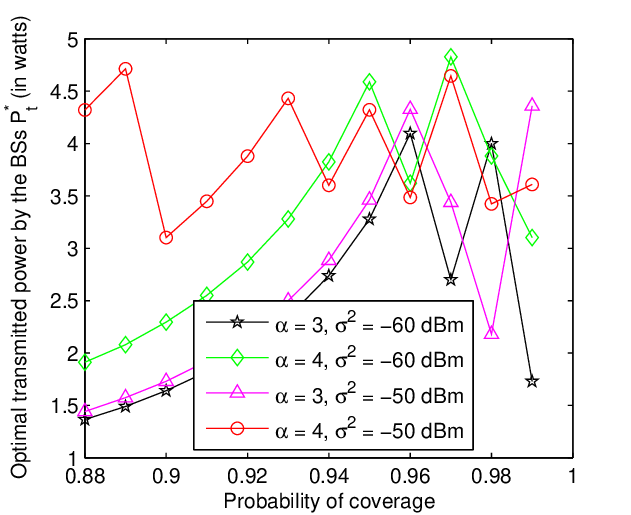}
	\caption{\small Optimal transmit power $P_t^*$ with coverage probability for different value of $\alpha$ and $\sigma^2$ over the square field.}
	\label{fig:optimal_Pt_square}
\end{figure}

If we examine Figs.~\ref{fig:optimal_Nb_circular}, \ref{fig:optimal_Nb_square}, \ref{fig:optimal_Pt_circular}, and \ref{fig:optimal_Pt_square}, Figs.~\ref{fig:optimal_Nb_circular}, \ref{fig:optimal_Nb_square} describe the optimal number of BSs $N_b^*$ and Figs.~\ref{fig:optimal_Pt_circular}, \ref{fig:optimal_Pt_square} depict the optimal transmit power $P_t^*$ with coverage probability over the two fields. In Fig.~\ref{fig:optimal_Nb_circular}, at $\alpha=3$, for the change of $\sigma^2$ from $-60$ dBm to $-50$ dBm, $N_b^*$ is same upto the $97\%$ coverage over the circular field. It can be explained using Fig.~\ref{fig:optimal_Pt_circular} where the optimal transmit power $P_t^*$ is higher for $\sigma^2=-50$ dBm which helps in compensation of the higher noise present in the channel. But, in case of the square field, at $\sigma^2=-50$ dBm, $N_b^*$ is increased (cf. Fig.~\ref{fig:optimal_Nb_square}) while keeping almost same transmit power (cf. Fig.~\ref{fig:optimal_Pt_square}) that compensate the noise by decreasing the cell size over the square field. We can also find that at around $97\%$ coverage, $N_b^*$ is increased abruptly but $P_t^*$ is decreased in large amount. Therefore, the abrupt changes do not take place in the total power consumption as shown in Figs.~\ref{fig:opt_pow_circular} and~\ref{fig:opt_pow_square}. 

\section{Summary}
This chapter describes the strategy for minimization of total power consumption while satisfying the desired coverage of the UEs to provide the minimum throughput over a wireless network. In order to achieve the goal, the deployment of BSs, their number, and transmit power are optimized in two scenarios: (i) when large number of UEs are present and (ii) when moderate UEs are distributed over a square or circular field. From the obtained numerical results, we find that the circular field is more energy-efficient than the square field in achieving the goal.

\section*{Appendix}
\addcontentsline{toc}{section}{Appendix}
\begin{figure}[!t]
	\sidecaption
	\includegraphics[scale=0.5]{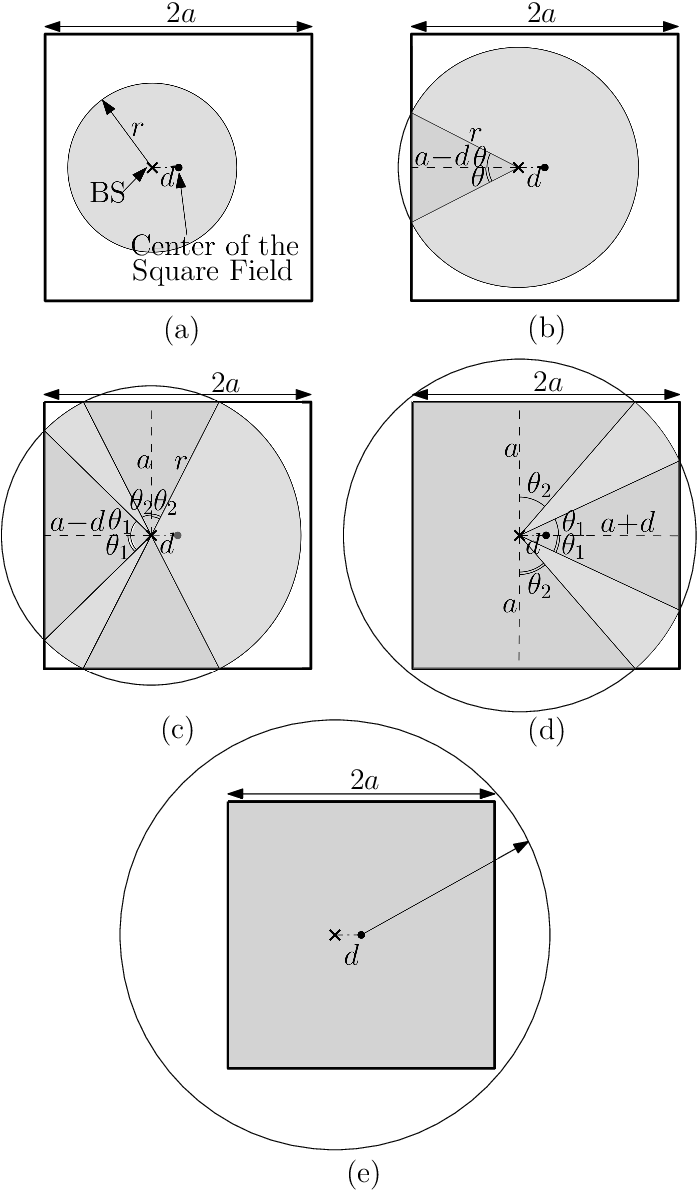}
	\caption{\small Distribution of distance of an UE from the BS located at distance $d$ from the center of the square field.}
	\label{fig:square_pdf_cdf}
\end{figure}
Here, we derive the distribution of distance of a UE which is located leftward at distance $d$ from the center of the square field as shown in Fig.~\ref{fig:square_pdf_cdf}. Probability that a UE lies at $\leq r$ distance from the BSs can be given by:
\begin{align}
	\mathrm{Pr}(R\leq r, d)=F(r,d)=\frac{Ar(C(r,d)\cap \Box)}{Ar(\Box)},
\end{align}
where $F(r,d)$ is the CDF of distance $R$ of a UE from the BS located at $d$ from the center of the square field, $C(r,d)$ is the circular field centered at the BS with radius $r$, $Ar(C(r,d)\cap \Box)$ is the area of intersection of the circular field and the square field, and $Ar(\Box)$ is the area of the square field. From Fig.~\ref{fig:square_pdf_cdf}, it can be observed that the intersection of the circular with square field changes with $r$. Moreover, the intersection also changes with $d$ from the center of the square field. Based on it, the CDF $F(r,d)$ for $d\in[0,\frac{a}{4}]$ can be determined as:

\begin{align}\label{eq36}
F(r, d)=
\begin{cases}
F_{11}=\frac{\pi r^2}{4a^2},   & \text{a}:\hspace{1mm} \text{$0 \leq r < a-d $}\\
F_{12}=\frac{1}{4a^2}\big[(a-d)\sqrt{r^2-(a-d)^2}\\ + \big(\pi - \cos^{-1}{\frac{(a-d)}{r}}\big)\frac{r^2}{2}\big], &\text{b}:\hspace{1mm} \text{$a-d \leq r < a$}\\
F_{13}=\frac{1}{4a^2}\big[(a-d)\sqrt{r^2-(a-d)^2}\\ + 2a\sqrt{r^2-a^2}\\ + \big(\pi - \cos^{-1}\frac{(a-d)}{r}\\ - 2\cos^{-1}\frac{a}{r}\big)r^2\big], &\text{c}:\hspace{1mm} \text{$a \leq r< a+d$}\\ F_{14}=\frac{1}{4a^2}\big[(a-d)\sqrt{r^2-(a-d)^2}\\+2a\sqrt{r^2-a^2}\\+(a+d)\sqrt{r^2-(a+d)^2}\\+\big(\pi-\cos^{-1}\frac{(a-d)}{r}-\\ \cos^{-1}\frac{(a+d)}{r}-2\cos^{-1}\frac{a}{r}\big)r^2\big], &\text{d}:\hspace{1mm} \text{$a+d\leq r$} \\ & \text{$<\sqrt{(a-d)^2+a^2}$}\\ F_{15}=\frac{1}{4a^2}\big[2a(a-d)+a\sqrt{r^2-a^2}\\+(a+d)\sqrt{r^2-(a+d)^2}\\ +\big(\pi/2-\cos^{-1}\frac{a}{r}\\-\cos^{-1}\frac{(a+d)}{r}\big)r^2\big], &\text{e}:\hspace{1mm} \text{$\sqrt{(a-d)^2+a^2} \leq r$}\\ &\text{$< \sqrt{(a+d)^2+a^2}$}\\
F_{16}=1, &\text{f}:\hspace{1mm} \text{$r \geq \sqrt{(a+d)^2 + a^2}$}.
\end{cases}
\end{align}
\begin{equation}\label{eq37}
f(r, d)=
\begin{cases}
f_{11}=\frac{\pi r}{2a^2},   &\text{a}:\hspace{1mm} \text{$0 \leq r < a-d $}\\
f_{12}=\frac{1}{2a^2}\big(\pi - \cos^{-1}{\frac{(a-d)}{r}}\big)r, &\text{b}:\hspace{1mm} \text{$a-d \leq r < a$}\\
f_{13}=\frac{1}{2a^2}\big(\pi - \cos^{-1}\frac{(a-d)}{r}\\ - 2\cos^{-1}\frac{a}{r}\big)r, &\text{c}:\hspace{1mm} \text{$a \leq r< a+d$}\\ f_{14}=\frac{1}{2a^2}\big(\pi-\cos^{-1}\frac{(a-d)}{r}-\\ \cos^{-1}\frac{(a+d)}{r}-2\cos^{-1}\frac{a}{r}\big)r, &\text{d}:\hspace{1mm} \text{$a+d\leq r$} \\ & \text{$<\sqrt{(a-d)^2+a^2}$}\\ f_{15}=\frac{1}{2a^2}\big(\pi/2-\cos^{-1}\frac{a}{r}\\-\cos^{-1}\frac{(a+d)}{r}\big)r, &\text{e}:\hspace{1mm} \text{$\sqrt{(a-d)^2+a^2} \leq r$}\\ &\text{$< \sqrt{(a+d)^2+a^2}$}\\
f_{16}=0, &\text{f}:\hspace{1mm} \text{$r \geq \sqrt{(a+d)^2 + a^2}$}.
\end{cases}
\end{equation}
Similarly, we can determine the CDF $F(r,d)$ and PDF $f(r,d)$ for other ranges of $d$. Now, using the same procedure, we determine the CDF $F(r,d)$ and PDF $f(r,d)$ of distance of a UE from the BS located at $d$ from the peak of the triangular field as shown in Fig.~\ref{fig:triangle_cdf_pdf}. For $d\in[0,\frac{a}{2}]$, the distribution can be determined as:

\begin{figure}[!t]
	\sidecaption
	\includegraphics[scale=0.35]{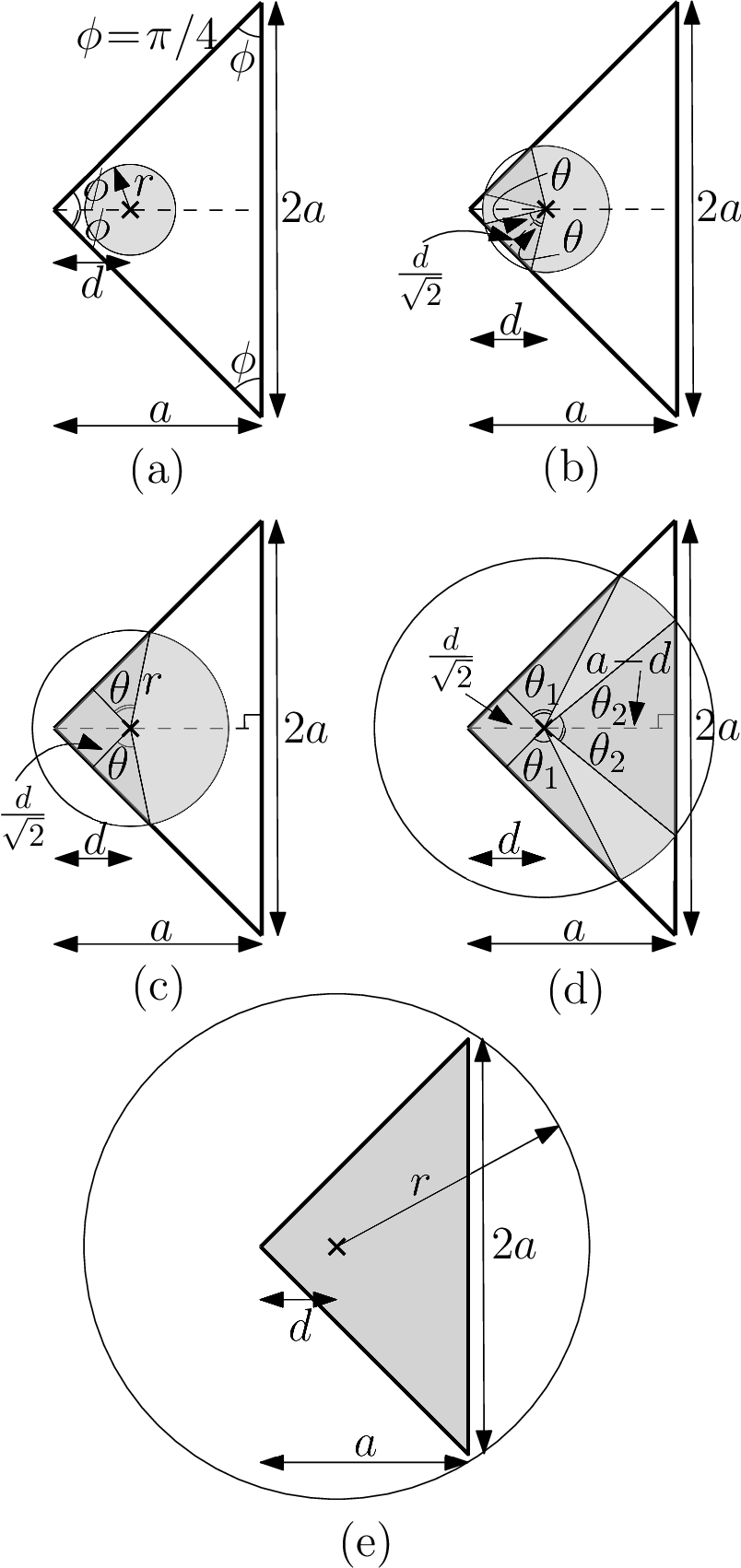}
	\caption{\small Distribution of distance of an UE from the BS located at distance $d$ from the peak of the triangular field.}
	\label{fig:triangle_cdf_pdf}
\end{figure}

\begin{equation}\label{eq41}
F(r, d)=
\begin{cases}
F_{21}=\frac{\pi r^2}{a^2},   &\text{p}: \text{$0 \leq r < \frac{d}{\sqrt{2}}$}\\
F_{22}=\frac{1}{a^2}\big[\big(\sqrt{2}d\sqrt{r^2-\frac{d^2}{2}} + (\pi\\ - 2\cos^{-1}{\frac{d}{r\sqrt{2}}}\big)r^2\big], &\text{q}: \text{$\frac{d}{\sqrt{2}} \leq r < d$}\\
F_{23}=\frac{1}{a^2}\big[\frac{d^2}{2}+\frac{d}{\sqrt{2}}\sqrt{r^2-\frac{d^2}{2}}\\+ \big(\frac{3\pi}{4} - \cos^{-1}\frac{d}{r\sqrt{2}}\big)r^2\big], & \text{r}: \text{$d \leq r< a-d$}\\ F_{24}=\frac{1}{a^2}\big[\frac{d^2}{2}+\frac{d}{\sqrt{2}}\sqrt{r^2-\frac{d^2}{2}}\\+(a-d)\sqrt{r^2-(a-d)^2}\\+\big(\frac{3\pi}{4}-\cos^{-1}\frac{d}{r\sqrt{2}}\\-\cos^{-1}\frac{(a-d)}{r}\big)r^2\big], & \text{s}: \text{$a-d\leq r$} \\ & \text{$<\sqrt{(a-d)^2+a^2}$}\\
F_{25}=1, & \text{t}: \text{$r \geq \sqrt{(a-d)^2 + a^2}$}.
\end{cases}
\end{equation}\\
\begin{equation}\label{eq42}
f(r, d)=
\begin{cases}
f_{21}=\frac{2\pi r}{a^2},   & \text{p}: \text{$0 \leq r < \frac{d}{\sqrt{2}}$}\\
f_{22}=\frac{2}{a^2}\big(\pi - 2\cos^{-1}{\frac{d}{r\sqrt{2}}}\big)r, & \text{q}: \text{$\frac{d}{\sqrt{2}} \leq r < d$}\\
f_{23}=\frac{2}{a^2}\big(\frac{3\pi}{4} - \cos^{-1}\frac{d}{r\sqrt{2}}\big)r, &\text{r}: \text{$d \leq r< a-d$}\\ f_{24}=\frac{2}{a^2}\big(\frac{3\pi}{4}-\cos^{-1}\frac{d}{r\sqrt{2}}\\-\cos^{-1}\frac{(a-d)}{r}\big)r, & \text{s}: \text{$a-d\leq r$} \\ &  \text{$<\sqrt{(a-d)^2+a^2}$}\\
f_{25}=0, & \text{t}: \text{$r \geq \sqrt{(a-d)^2 + a^2}$}.
\end{cases}
\end{equation}\\
Similarly, we can compute the distribution for other ranges of $d$.

\end{document}